\documentclass[
 reprint,amsmath,amssymb,aps,fleqn,showpacs
]{revtex4-1}
\usepackage{graphicx}
\usepackage{dcolumn}
\usepackage{bm}
\begin{document}


\title{Moment fluid equations for ions in weakly-ionized plasma}

\author{I.$\,$L.$\,$Semenov}
\email{Igor.Semenov@.dlr.de}
\affiliation{Forschungsgruppe Komplexe Plasmen, Deutsches Zentrum f\"{u}r Luft- und Raumfahrt, Oberpfaffenhofen,
Germany}

\date{\today}

\begin{abstract}
\noindent
A new one-dimensional fluid model for ions in weakly-ionized plasma is proposed.~The model differs from the existing ones in two aspects. First, a more accurate approximation of the collision terms in the fluid equations is suggested. For this purpose, the results obtained using the Monte-Carlo kinetic model of the ion swarm experiments are considered.
Second, the ion energy equation is taken into account.~The fluid equations are closed using a simple model of the ion velocity distribution function.
The accuracy of the fluid model is examined by comparing with the results of particle-in-cell/Monte Carlo simulations. In particular, several test problems are considered using a parallel plate model of the capacitively coupled radio-frequency discharge. It is shown that the results obtained using the proposed fluid model are in good agreement with those obtained from the simulations over a wide range of discharge conditions. An approximation of the ion velocity distribution function for the problem under consideration is also discussed.
\end{abstract}



\maketitle

\section{Introduction}
Mathematical and numerical modeling of transport processes in weakly-ionized plasma plays an important role for better understanding the basic properties of low-pressure discharges and further development of plasma assisted technologies. As it is known, there are two main types of models for describing transport phenomena in partially-ionized plasma: 
kinetic and fluid (continuum) models. A general description of these approaches can be found in many review papers and textbooks \cite{golant1980fundamentals,zhdanov2002transport,kim2005particle}.
\\ \indent 
For example, kinetic models are more preferable for describing the electron component in low-pressure discharges because non-equilibrium and non-local kinetic effects have a substantial influence on the electron distribution function \cite{tsendin1995electron,hagelaar2005solving,kolobov2006simulation}.
Fluid equations for electrons can be reliably applied only at sufficiently high pressures and low electric fields provided that the transport coefficients are calculated using the kinetic approach \cite{hagelaar2005solving}. In contrast to the electrons the ion component in low-pressure discharges is more often described by means of the fluid equations \cite{richards1987continuum,meyyappan1990glow,gogolides1992continuum1,
gogolides1992continuum2,passchier1993two,nitschke1994comparison,
boeuf1995two,becker2016advanced,chen2016numerical}. Despite the fact that these equations are not always based on an explicitly stated kinetic model they have been shown to give reasonable results under certain conditions. Moreover, such models are widely used to analyze the presheath-sheath transition in weakly-ionized plasma \cite{godyak1982modified,sternberg2003patching,
sternberg2007bohm,brinkmann2011plasma}. Nevertheless, two points can be made regarding the applicability and accuracy of the fluid equations for ions.
\\ \indent
At first, it should be noted that these equations are generally formulated without considering the ion energy balance. In order to close the model, the pressure term in the ion momentum equation is neglected or assumed to be equal to that of the local Maxwellian distribution with the gas temperature. Since this assumption is not well justified in the presheath-sheath region, it 
might affect the accuracy of the model and must be examined carefully  \cite{surendra1993moment}. Furthermore, if the fluid model does not reproduce the effects associated with the ion energy transfer, it has a limited ability to describe, at least qualitatively, the ion velocity distribution.~On the other hand, this might be important for theoretical studies of ion flows in weakly-ionized plasma.
\\ \indent
The second point to note is the approximation used for the velocity moments of the ion-neutral collision integral. A common approach in this case is to apply the well-known exchange relations \cite{golant1980fundamentals,zhdanov2002transport} with an effective collision frequency. The latter is usually evaluated as a function of the ion mean velocity using the data from the ion swarm experiments \cite{brinkmann2011plasma,becker2016advanced}. It is worth noting that the experimental data are available only in a limited range of the electric fields \cite{ellis1976transport,phelps1991cross} and cannot be used to calculate the effective collision frequency in the entire range of the ion drift velocities. For this reason, it is more appropriate to use theoretical or numerical models of the ion swarm experiments in this case.
\\ \indent
In the present work we propose a one-dimensional fluid model for ions which partially overcomes the limitations discussed above. To address these limitations the following steps have been implemented. First, we consider the Monte-Carlo (MC) kinetic model of the ion swarm experiments to define the approximations of the collision terms in the fluid equations. Second, the ion energy equation is taken into account. The fluid equations are closed by applying a simple model of the ion distribution function (IDF).
The accuracy of the proposed fluid model is examined by comparing the predicted results with those obtained using the particle-in-cell simulation method combined with the Monte-Carlo collision model (PIC-MCC approach) \cite{vahedi1995monte,verboncoeur2005particle,donko2011particle}. In particular, a parallel plate model of the capacitively coupled radio-frequency (CCRF) discharge in argon and helium is considered. The results are presented for several test problems including the benchmarks \cite{turner2013simulation} and experimental situations \cite{godyak1990abnormally,godyak1992measurement}.
On the basis of the obtained results, a possible approximation of the IDF for the problem under consideration is discussed.
\\ \indent
The structure of the paper is as follows. 
In Section~\ref{Sec: KinModel} the Monte-Carlo kinetic model of the ion swarm experiments is described and the approximation for the moments of the ion-neutral collision integral is discussed. 
In Section~\ref{Sec: FluidModel} the fluid equations for ions are presented. The accuracy of the model is analyzed and discussed in Section~\ref{Sec: Results}. The conclusions are drawn in Section~\ref{Sec: Concl}.
\section{Kinetic model of the ion swarm experiments}
\label{Sec: KinModel}
\begin{figure}[!t]
\includegraphics[width=0.5\textwidth]{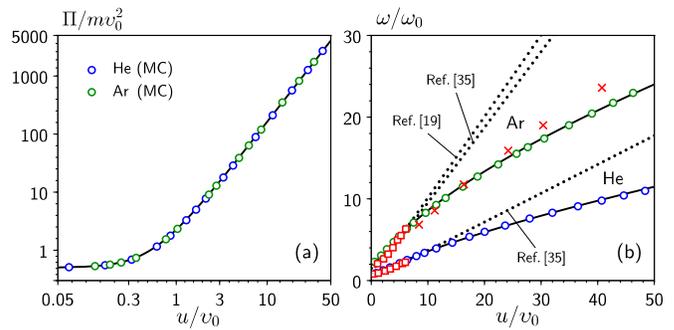}
\caption{\label{Fig: swarm} (a) The gas-dynamic pressure as a function of the ion drift velocity (a logarithmic scale
is used for both axes).~Circles show the results obtained from the MC simulations for helium (blue) and argon (green).~The solid line shows the approximation given by Eq.~(\ref{P}). (b) The effective collision frequency as a function of the ion drift velocity.
Circles show the results obtained from the MC simulations for helium (blue) and argon (green).~Solid lines show the approximations given by Eq.~(\ref{omega_fit}).~Squares and crosses show the results obtained using the experimental data of Ref.~\cite{ellis1976transport} and data of Phelps \cite{phelps1991cross}, respectively.~Dotted lines show the approximations proposed in Refs.~\cite{brinkmann2011plasma,khrapak2013practical}.}
\end{figure}

\begin{figure}[!t]
\includegraphics[width=0.5\textwidth]{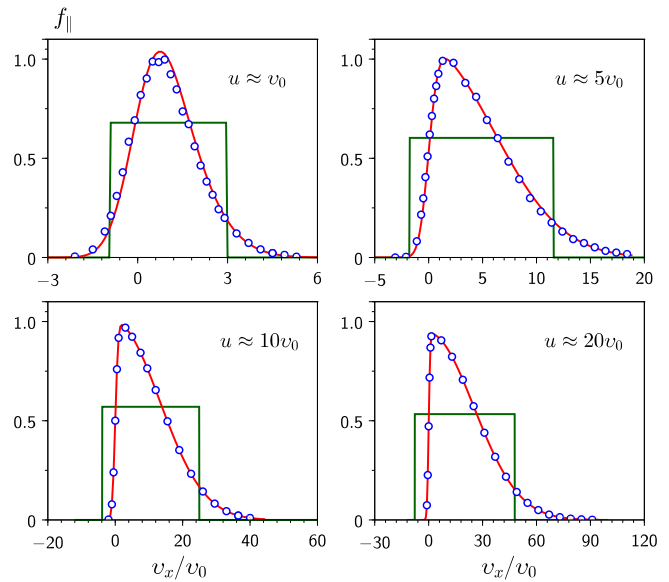}
\caption{\label{Fig: swarm_df} The parallel IDF ($f_{\parallel}$) for  drift flows. Circles show the IDF obtained from the MC simulations. Solid lines show the IDF calculated using the model of Ref.~\cite{lampe2012ion} (red line) and model~(\ref{f_band}) (green line).~For simplicity, all results are normalized by the maximum value of the IDF obtained from the MC simulations.} 
\end{figure}

Let us first discuss the Monte-Carlo (MC) kinetic model of the ion swarm experiments. Namely, we consider a flow of ions in weakly-ionized plasma under the influence of a uniform and constant electric field. It is assumed that the ions are singly charged and experience elastic and charge exchange collisions with gas atoms. In order to find the steady-state IDF for the problem under consideration, we have used the MC simulation approach similar to that described in Ref.~\cite{lampe2012ion}. The ion-neutral collisions have been treated using the MC model presented in Ref.~\cite{vahedi1995monte}. The simulations have been performed 
for argon and helium plasma in a wide range of  
$E/n_{a}$, where $E$ is the absolute value of the electric field and $n_{a}$ is the number density of the gas atoms.
The gas temperature, $T_{a}$, has been set to 300K.
\\ \indent
In the case of helium, the cross-sections for ion-neutral collisions were defined using the dataset of Phelps available in the database LXCat \cite{LXCat}. In the case of argon, the cross-section for elastic ion-neutral scattering was defined using the approximation proposed by Phelps in Ref.~\cite{phelps1994application}. The cross-section for charge exchange collisions in argon  was evaluated using the formula presented by Devoto in Ref.~\cite{devoto1967transport}. This approximation was derived to fit the experimental data of Ziegler \cite{ziegler1953wirkungsquerschnitt} and 
Cramer \cite{cramer1959elastic}. It should be noted that the approximation of Devoto differs from that proposed by Phelps in Ref.~\cite{phelps1994application} for the charge exchange cross-section.
One can show, however, that the approximation of Phelps, when used in PIC-MCC simulations of CCRF discharges, causes
noticeable deviations from the well-known experimental data of Godyak \citep{godyak1990abnormally}. On the other hand, the PIC-MCC simulations employing the formula of Devoto give reasonable agreement with the data of Godyak (see Sec.~\ref{Sec: Results}). For this reason, the results presented below for argon  were obtained using the approximation of Devoto.\\ \indent
Let us now consider the results of numerical simulations performed using the MC model of the ion swarm experiments. The first quantity of interest is the ion gas-dynamic pressure
\begin{align*}
\Pi=\int m \upsilon_{x}^{2} \, f d \vec{\upsilon},
\end{align*}
where $m$ is the ion mass, 
$\vec{\upsilon}$ is the ion velocity vector, $\upsilon_{x}$ is the projection of $\vec{\upsilon}$ on the axis directed along the electric field vector and $f$ is the IDF normalized as 
$\int f d \vec{\upsilon}=1$. The values of $\Pi$ for argon and helium are plotted in Figure~\ref{Fig: swarm}(a) as a function of the ion drift velocity, $u$, normalized by the atom thermal velocity
$\upsilon_{0}=\sqrt{2kT_{a}/m}$.\\ \indent
As it can be seen in Fig.~\ref{Fig: swarm}(a) the results obtained for both gases lie approximately on the same curve which can be fitted taking into account the following considerations. It is known that the IDF for ion drift flows can be found analytically by considering two simple models of the ion-neutral collision integral \cite{lampe2012ion}. Following the results of Ref.~\cite{lampe2012ion}, one can show that the analytical solution gives
$\Pi=kT_{a}+2 \, m u^{2}$
for $u \lesssim \upsilon_{0}$ and
$\Pi=(\pi/2) \,m u^{2}$
for $u \gtrsim \upsilon_{0}$. Keeping in mind these estimates, the following approximation of the numerical results has been proposed:
\begin{align}
& \Pi=kT_{a}+\alpha(\xi) \, mu^{2}, \nonumber\\
\label{P}
& \alpha(\xi)=1.038(\pi/2)(1-e^{-\xi})+2e^{-\xi},
\end{align}
where $\xi=u / \upsilon_{0}$. Eq.~(\ref{P}) interpolates between the limiting values of $\Pi$ known from the analytical solution. 
In the regime $u \gg \upsilon_{0}$ the coefficient before $mu^{2}$ is corrected to obtain better agreement with the numerical results. 
The function $\Pi(u)$ given by Eq.~(\ref{P}) is plotted in Fig.~\ref{Fig: swarm}(a). It can be seen that this approximation agrees well with the results of numerical simulations.
\\ \indent
The second parameter of interest is the effective collision frequency, 
$\omega$, mentioned before. According to the common definition, we adopt $\omega=eE/mu$. The values of 
$\omega$ obtained from the numerical simulations are shown in Figure~\ref{Fig: swarm}(b) as a function of $u/\upsilon_{0}$. Note that $\omega$  in 
Fig.~\ref{Fig: swarm}(b) is normalized by 
$\omega_{0}=n_{a} \sigma_{0} \upsilon_{0}$, where 
$\sigma_{0}=10^{-18}\,$m$^{-2}$. As a result
$\omega/\omega_{0}$ depends on $E/n_{a}$ and, consequently, can be considered as a function of the drift velocity. For comparison, we show in Fig.~\ref{Fig: swarm}(b) the results obtained using the experimental data of Ref.~\cite{ellis1976transport} and those given by several models published previously. In particular, we show the approximations proposed for argon and helium in Ref.~\cite{khrapak2013practical}  by considering the experimental data of Frost \cite{frost1957effect}. In the case of argon, we also show the 
approximation presented in Ref.~\cite{brinkmann2011plasma} and 
the results obtained using the data of Phelps \cite{phelps1991cross} which are based on the theoretical model  of Ref.~\cite{lawler1985equilibration}.
As it can be seen from Fig.~\ref{Fig: swarm}(b), accurate approximation of the function $\omega(u)$ is of importance. For example, the expressions  presented in Refs.~\cite{brinkmann2011plasma,khrapak2013practical} lead to noticeable deviations from the MC simulation results in the regime $u \gg \upsilon_{0}$.
\\ \indent
Moreover, the numerical results obtained in the present work can be easily fitted by the following expression:
\begin{equation}
\label{omega_fit}
\omega / \omega_{0}= \left \{
\begin{array}{ll}
\nu_{1} \, e^{\,w_{1} \xi}  & 
\quad \xi \le \xi_{*} \\
\nu_{2} \, \xi^{\,w_{2}} & 
\quad \xi > \xi_{*}
\end{array}
\right.,
\end{equation}
where $\xi=u/\upsilon_{0}$, $w_{1}=w_{2}/\xi_{*}$ and $\xi_{*}$ is the solution of the equation 
$w_{2} \ln \xi_{*}=w_{2}+\ln \nu_{1}-\ln \nu_{2}$. 
Eq.~(\ref{omega_fit}) provides a smooth fit to the numerical results in a wide range of the ion drift velocities. 
The constants in Eq.~(\ref{omega_fit}) are 
$\xi_{*} \approx 2.81$, $\nu_{1}=2.00$, $\nu_{2}=1.96$, 
$w_{1} \approx 0.23$, $w_{2}=0.64$ for argon and 
$\xi_{*} \approx 3.11$, $\nu_{1}=0.75$, $\nu_{2}=0.68$, 
$w_{1} \approx 0.23$, $w_{2}=0.72$ for helium. The fitting formula~(\ref{omega_fit}) has been found to agree well with the results of the MC simulations (see Fig.~\ref{Fig: swarm}(b)).\\ \indent
The third point of interest is the approximation of the IDF for the problem under consideration. As it was demonstrated in Ref.~\cite{lampe2012ion}, the IDF for drift flows can be approximated with good accuracy using a simple model based on the assumption of constant cross-section for ion-neutral collisions.
Following Ref.~\cite{lampe2012ion}, the IDF is written as $f(\vec{\upsilon})=f_{\parallel}(\upsilon_{x})f_{\perp}(\upsilon_{\perp})$, where 
$\upsilon_{\perp}=\sqrt{\vec{\upsilon}^{\,2}-\upsilon_{x}^{2}}$, $f_{\perp}=(\pi \upsilon_{0}^{2})^{-1}\,\exp(-\upsilon_{\perp}^{2}/\upsilon_{0}^{2})$
and $f_{\parallel}$ is found by solving a simple
ordinary differential equation.
Note that this model neglects the effect of elastic ion-neutral collisions on the transverse distribution $f_{\perp}$.
Strictly speaking, this assumption is justified only for the case when the charge-exchange collisions are dominant (e.g., for noble gases).
The comparison between the model of Ref.~\cite{lampe2012ion} and the  results of our MC simulations is shown in Figure~\ref{Fig: swarm_df} (for brevity, only the case of argon is presented). It can be seen that the model of 
Ref.~\cite{lampe2012ion} agrees well with the results of the MC simulations in a wide range of the ion drift velocities.
In addition, we show in Fig.~\ref{Fig: swarm_df} the most simple and general model of the IDF formulated as
\begin{equation}
\label{f_band}
f_{\parallel}= \left \lbrace 
\begin{array}{ll}
1 /(2 \upsilon_{c}) &  \quad  
u-\upsilon_{c}<\upsilon_{x} < u+\upsilon_{c}\\
0 & \quad \mathrm{otherwise}
\end{array}
\right.,
\end{equation}
where $\upsilon_{c}$  is defined by the value of the gas-dynamic pressure.
In the present case we have used Eq.~(\ref{P}) to evaluate $\upsilon_{c}$ for the IDF shown  in Fig.~\ref{Fig: swarm_df}.
As one can see, despite its simplicity, the model~(\ref{f_band}) is able to give a reasonable
estimate of the velocity range where the IDF is essentially non-zero.
\\ \indent
Finally, let us discuss an approximation for the velocity moments of the ion-neutral collision integral. In fact, using Eqs.~(\ref{P}) and (\ref{omega_fit}), one can easily formulate simple moment equations
which reproduce the kinetic solution for $u$ and $\Pi$. These equations are
\begin{align}
& eE - \omega \, m  u=0, \nonumber \\
\label{swarm_moment}
& 2euE+\omega \,\delta  \left(  kT_{a}-\Pi \right)=0,
\end{align}
where $\delta=2/\alpha(u)$.
Equations~(\ref{swarm_moment}) approximate the velocity moments
of the Boltzmann equation corresponding to the weights 
$m\upsilon_{x}$
and $m\upsilon_{x}^{2}$, respectively. It should be noted that similar moment equations can be obtained by considering the Bhatnagar-Gross-Krook model  of the collision integral \cite{lampe2012ion} with the collision frequency $\omega$. The only difference in Eqs.~(\ref{swarm_moment}) is that we have introduced the correction coefficient $\delta$ to reproduce the kinetic solution for the gas-dynamic pressure. 
In conclusion, we suggest to use moment equations~(\ref{swarm_moment}) as a basis for developing more detailed fluid models.
\section{Fluid equations for ions}
\label{Sec: FluidModel}
Using the results of Sec.~\ref{Sec: KinModel}, we propose a new fluid model for ions in weakly-ionized plasma. For simplicity a one-dimensional flow along the $x$-axis is considered. Let us introduce the variables $p$ and $q$ using the following relations:
\begin{equation*}
\Pi = mu^{2}+p, \quad
\int m \upsilon_{x}^{3} \, f d \vec{\upsilon} = mu^{3}+qu.
\end{equation*}
Then, the governing equations of the model are written as
follows:
\begin{equation}
\label{gd_eq}
\frac{\partial \mathbf{U}}{\partial t}+
\frac{\partial \mathbf{F}}{\partial x}=\mathbf{H},
\end{equation}
where $\mathbf{U}=[n, mnu, n \Pi ]^{\mathrm{T}}$ and
\begin{align*}
&
\mathbf{F}=\left[ 
\begin{array}{c}
nu \\ 
n\Pi \\
(1-\gamma)mnu^{3}+\gamma n \Pi u
\end{array}
\right],\\
&
\mathbf{H}=\left[
\begin{array}{c}
G\\
neE-\omega mnu \\
2nueE+\omega \, \delta n (kT_{a}-\Pi)+G k T_{a}
\end{array}
\right].
\end{align*}
Here $n$ is the ion number density, $G$ is the ionization source and the notation 
$\gamma=q/p$ is introduced. 
Eqs.~(\ref{gd_eq}) are obtained by considering the velocity moments of the Boltzmann equation with the weights 1, $m\upsilon_{x}$ and
$m\upsilon_{x}^{2}$, respectively. The collision terms in Eqs.~(\ref{gd_eq}) are approximated using the expressions employed in Eqs.~(\ref{swarm_moment}) of Sec.~\ref{Sec: KinModel}.
In order to close Eqs.~(\ref{gd_eq}), a certain approximation for $\gamma$ has to be applied. In the present work we evaluate $\gamma$ using the IDF considered in Sec.~\ref{Sec: KinModel} with $f_{\parallel}(\upsilon_{x})$ given by Eq.~(\ref{f_band}). In this case we get
\begin{align}
p=m\upsilon_{c}^{2}/3, \quad
q=m \upsilon_{c}^{2}, \quad
\gamma=3.
\end{align}
\indent
Although the model~(\ref{f_band}) is very rough, it can provide qualitative description of different ion velocity distributions
expected in low-pressure discharges. For example, at 
$\upsilon_{c} \lesssim \upsilon_{0}$ and
$|u| \gg \upsilon_{0}$ the model~(\ref{f_band}) represents a high-energy ion beam. Such distributions can be expected to occur when the convective terms play a dominant role in the momentum and energy balance. For instance, the model~(\ref{f_band}) was used in Refs.~\cite{stangeby1984plasma,benilov1995model} for studying the sheath region.
\\ \indent
In the opposite limit, when the collision terms are dominant, the steady-state solution of Eqs.~(\ref{gd_eq}) for $u$ and $\Pi$ tends to that given by Eqs.~(\ref{swarm_moment}). In this case, the IDF is expected to be close to the IDF for drift flows evaluated at the local mean velocity. Under such conditions, the model~(\ref{f_band}) predicts at least the velocity range where the IDF is essentially non-zero (see Sec.~\ref{Sec: KinModel}).
The value of $\gamma$ evaluated using the IDF for drift flows is higher than that given by the model~(\ref{f_band}). For example, using the well-known analytical result \cite{lampe2012ion}, one can show that the IDF for drift flows gives 
$\gamma=(\pi-1)/(\pi/2-1) \approx 3.75$ in the regime
$u \gtrsim \upsilon_{0}$. On the other hand, the variations of 
$\gamma$ do not affect significantly the accuracy of the fluid model when the ion flow is collision-dominated.
\\ \indent
When $\gamma$ is constant,  Eqs.~(\ref{gd_eq}) are of hyperbolic type. Therefore, these equations can be solved using the well developed numerical methods for hyperbolic problems \cite{leveque2002finite,toro2013riemann}.
In particular, in the present work we have applied an explicit flux-vector splitting scheme described in 
Ref.~\cite{steger1981flux}. 
The details of the numerical method are given in the Appendix.
It is worth noting that the ion flows in low-pressure discharges are expected to be described by smooth solutions of Eqs.~(\ref{gd_eq}) (i.e., without shock waves and contact discontinuities). From this point of view, the flux-vector splitting scheme seems to be a reasonable choice, because
it is easy to implement and provides sufficient accuracy
for smooth solutions.
\section{Results and Discussion}
\label{Sec: Results}
\begin{figure*}[!t]
\includegraphics[width=0.8\textwidth]{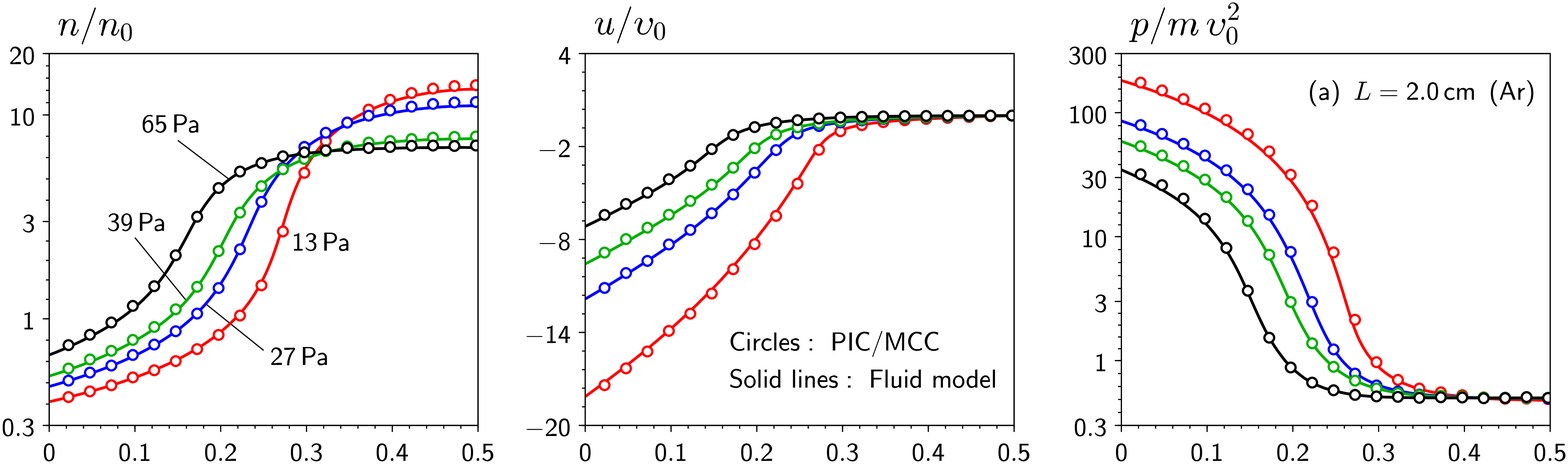}
\includegraphics[width=0.8\textwidth]{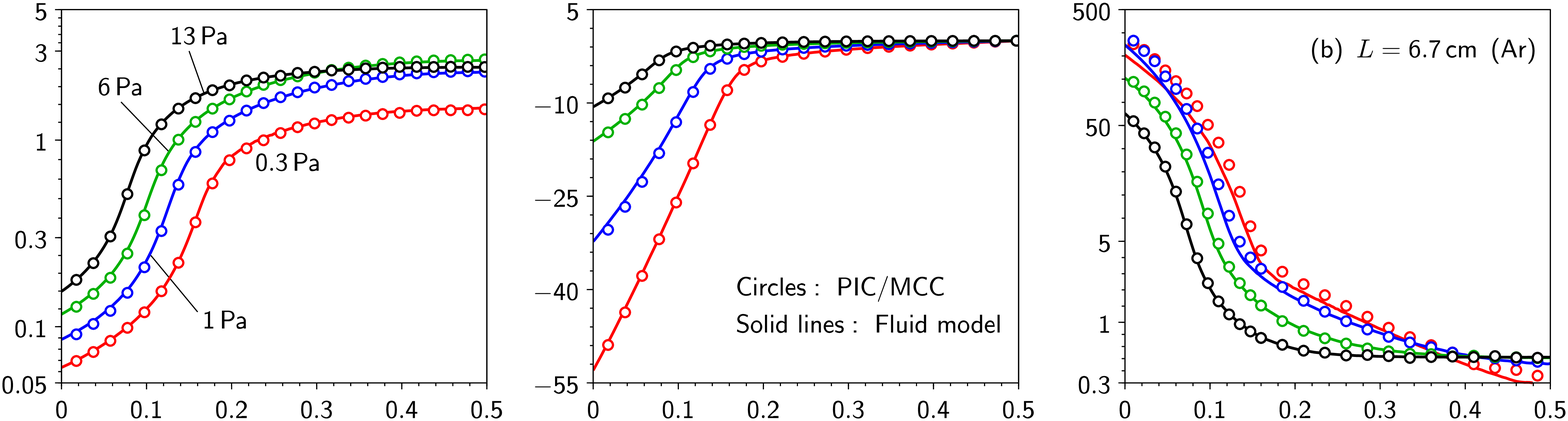}
\includegraphics[width=0.8\textwidth]{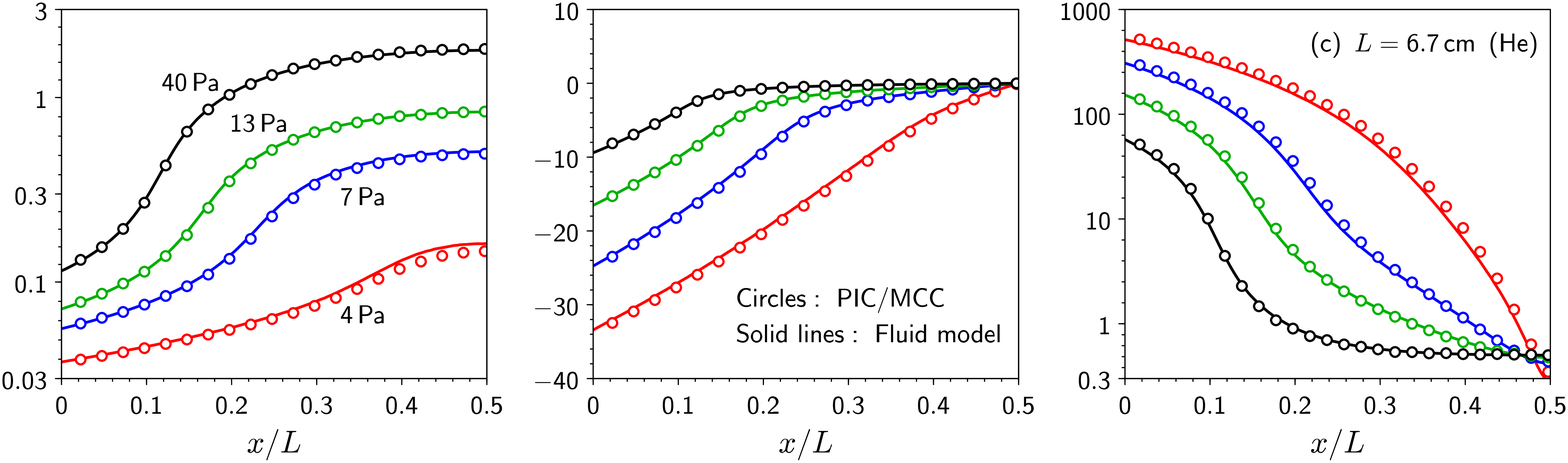}
\caption{\label{fig: param} Distributions of the ion number density
($n$), mean velocity ($u$) and internal energy ($p$) for different discharge conditions (here $n_{0}=10^{15} \, \mathrm{m}^{-3}$ and  a logarithmic scale
is used for $n$ and $p$). Solid lines show the results obtained using the fluid model described in Sec.~\ref{Sec: FluidModel}. Circles show the results obtained from the PIC-MCC simulations. Figure (a) shows the results for the first test problem (argon, $L=2\,$cm). The corresponding gas pressures and voltage amplitudes are: 13$\,$Pa, 260$\,$V (red); 27$\,$Pa, 212$\,$V (blue); 39$\,$Pa, 195$\,$V (green); 
65$\,$Pa, 160$\,$V(black).  Figure (b) shows the results for the second test problem (argon, $L=6.7\,$cm). The corresponding gas pressures and voltage amplitudes are: 0.3$\,$Pa, 200$\,$V (red); 1$\,$Pa, 160$\,$V (blue); 6$\,$Pa, 130$\,$V (green), 13$\,$Pa, 100$\,$V (black). 
Figure (c) shows the results for the third test problem (helium, $L=6.7\,$cm). The corresponding gas pressures and voltage amplitudes are: 4$\,$Pa, 450$\,$V (red); 7$\,$Pa, 300$\,$V (blue); 13$\,$Pa, 200$\,$V (green); 40$\,$Pa, 150$\,$V (black).  }
\end{figure*}
\begin{figure*}[!t]
\includegraphics[width=0.9\textwidth]{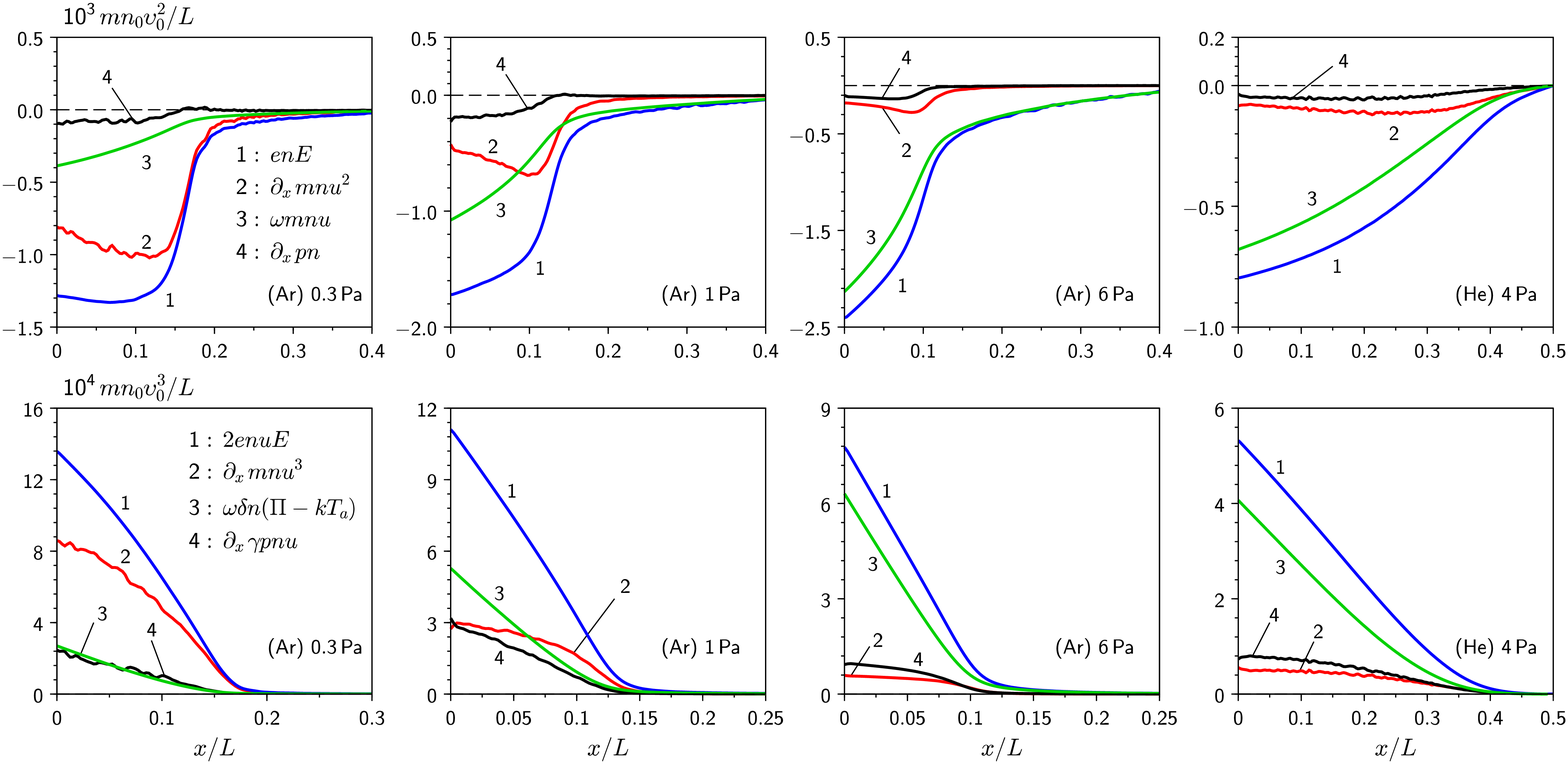}
\caption{\label{fig: terms_Ar} Distributions of different terms in the ion momentum and energy equations (upper and lower rows, respectively). Here $n_{0}=10^{15} \, \mathrm{m}^{-3}$. The results are presented for the second test problem (argon, $L=6.7\,$cm) at the gas pressures 0.3, 1, 6$\,$Pa and third test problem (helium, $L=6.7\,$cm) at the gas pressure of 4$\,$Pa.}
\end{figure*}
\begin{figure*}[!t]
\includegraphics[width=0.8\textwidth]{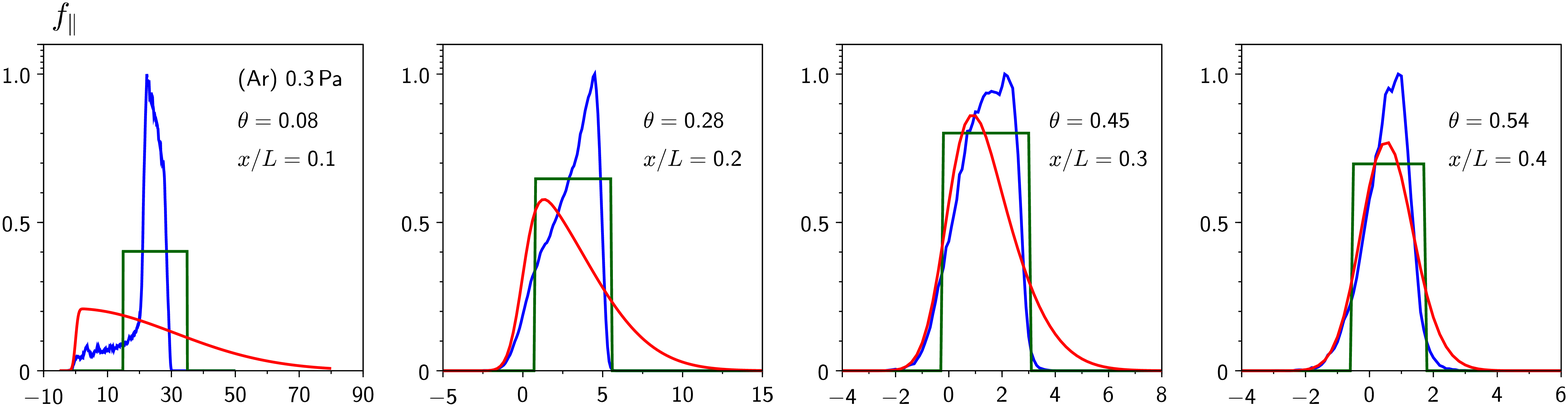}
\includegraphics[width=0.8\textwidth]{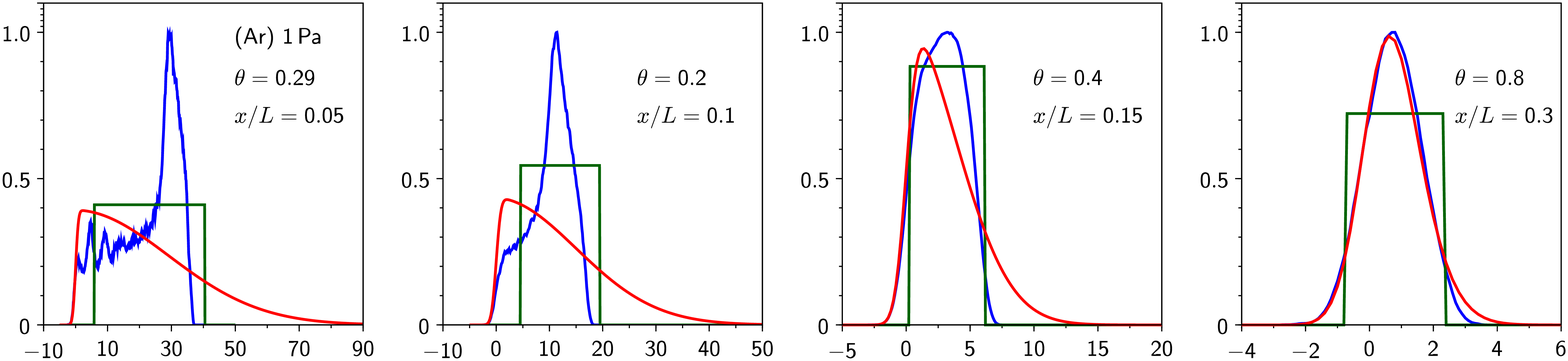}
\includegraphics[width=0.8\textwidth]{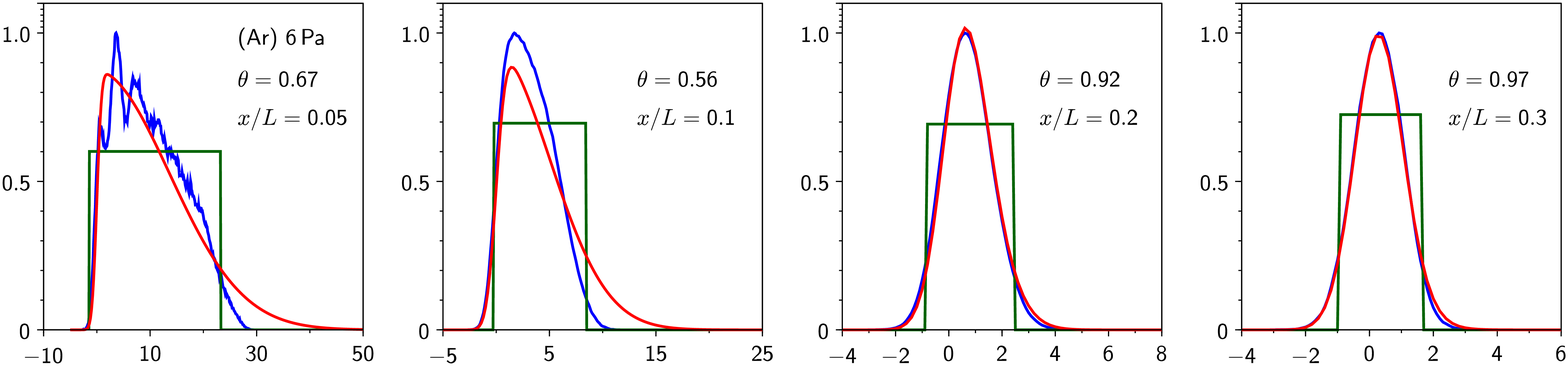}
\includegraphics[width=0.8\textwidth]{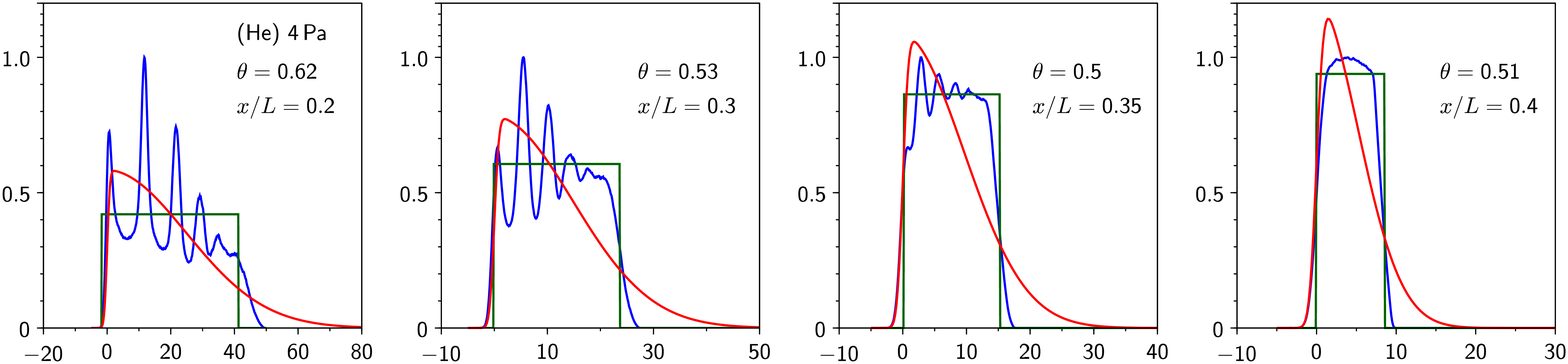}
\includegraphics[width=0.8\textwidth]{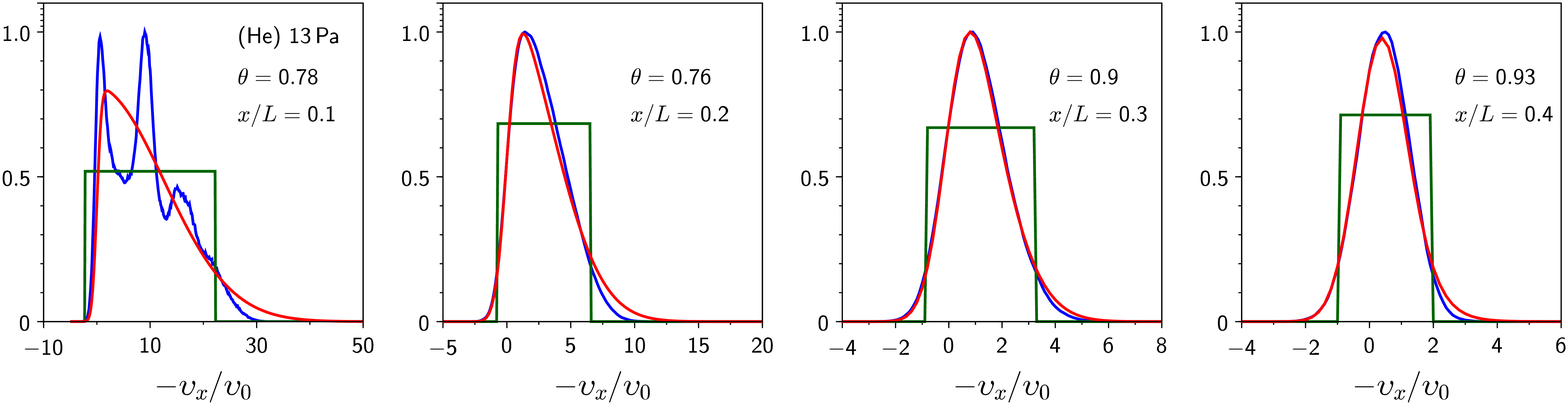}
\caption{\label{fig: df_dis} The parallel IDF ($f_{\parallel}$) at different positions in the discharge. The results are presented for the second test problem (argon, $L=6.7\,$cm) at the gas pressures 0.3, 1, 6$\,$Pa and third test problem (helium, $L=6.7\,$cm) at the gas pressures 4 and 13$\,$Pa. Blue lines show the IDF obtained from the PIC-MCC simulations. Red lines show the IDF for drift flows calculated using the model of Ref.~\cite{lampe2012ion}. Green lines show the IDF given by the model~(\ref{f_band}).~For simplicity, all results are normalized by the maximum value of the IDF obtained from the PIC-MCC simulations.}
\end{figure*}
In this section, the accuracy of the fluid equations proposed in Sec.~\ref{Sec: FluidModel} is analyzed by comparing with the simulations based on the PIC-MCC approach. In particular, we consider a parallel plate model of the CCRF discharge used previously in many works, e.g., in Refs.~\cite{vahedi1993capacitive1,
vahedi1993capacitive2, kim2005PIC, turner2013simulation}. 
Our PIC-MCC code is based on the conventional technique described in Refs.~\cite{donko2011particle,verboncoeur2005particle, vahedi1995monte}. The code was tested to reproduce the benchmark of Ref.~\cite{turner2013simulation} for helium and the 
results of Refs.~\cite{vahedi1993capacitive2,kim2005PIC} for argon.
\\ \indent
The collision model employed in our PIC-MCC simulations was defined as follows.~For ion-neutral collisions we used the same model as in 
Sec~\ref{Sec: KinModel}. The model of electron-neutral collisions included elastic scattering, ionization and excitation processes.~In the case of argon,
the cross-section of elastic electron-neutral collisions was defined using the dataset of Hayashi available in the database LXCat \cite{LXCat}.~The ionization cross-section was defined using the data of Smith \cite{smith1930ionization}. The excitation process was modeled using an effective level with the threshold 11.55~eV. The excitation cross-section was defined by summing the combined cross-sections for S, P and D levels taken from Biagi-v7.1 database \citep{LXCat}. The differential cross-section for electron-neutral collisions was defined using the model proposed in Ref.~\cite{okhrimovskyy2002electron}.~In the case of helium,
we used the same data for electron-neutral collisions as those described in Ref.~\cite{turner2013simulation}\\ \indent
To examine the accuracy of the fluid equations~(\ref{gd_eq}), we  compared the results of the kinetic PIC-MCC simulations with those obtained using a hybrid fluid-kinetic model. The fluid-kinetic model combined the fluid equations for ions and the PIC-MCC model for electrons. The fluid equations for ions were solved numerically by means of the flux-vector splitting scheme described in the Appendix. 
The ionization source in Eqs.~(\ref{gd_eq}) was evaluated using the electron velocity distribution function obtained from the PIC-MCC model.~The comparison was performed for three test problems.~For simplicity, the voltage boundary condition was used in all cases. 
\\ \indent
The discharge parameters for the first test problem were chosen
to represent the experimental conditions for argon used by Godyak and Piejak in Ref.~\cite{godyak1990abnormally}. Namely, the discharge length, $L$, was set to 2$\,$cm and the discharge frequency was set to 13.56$\,$MHz. The amplitude of the voltage was taken from the simulations performed  with the current boundary condition at the current density of 2.65\,$\mathrm{mA\,cm}^{-2}$. The gas pressure was varied in the range from 13 to 65$\,$Pa. Numerical parameters were chosen to reproduce the results of Ref.~\cite{kim2005PIC}. In particular, the number of cells was set to 400, the number of time steps within one RF period was set to 2000 and the final number of simulation particles for each component was of the order of $2 \times 10^{5}$.
The second test problem was chosen to represent the experiments of Godyak {\it et al.} presented in Ref.~\cite{godyak1992measurement} for argon at $L=6.7\,$cm.
The discharge frequency was left unchanged and the amplitude of the voltage was taken from the simulations performed with the current boundary condition at the current density of 1\,$\mathrm{mA\,cm}^{-2}$. The gas pressure was varied in the range from 0.3 to 13$\,$Pa. The numerical parameters were chosen to be the same as for the first test problem.
The third test problem was based on the benchmark for helium published in Ref.~\cite{turner2013simulation}. Namely, we performed simulations for the test cases 1-3 of Ref.~\cite{turner2013simulation} and a number of simulations with similar discharge conditions.\\ \indent
Let us now discuss the obtained results. In Figure~\ref{fig: param} we present the distributions of the period averaged ion number density ($n$), mean velocity ($u$) and internal energy ($p$) obtained using the kinetic PIC-MCC simulations and the fluid model
for the test problems described above. The distributions are presented for the left half part of the discharge, because the ion flow is symmetric with respect to the discharge center plane.
The corresponding discharge parameters are given in the caption to Fig.~\ref{fig: param}. As it can be seen from Fig.~\ref{fig: param}, the results obtained using the fluid model are in good agreement with the results of PIC-MCC simulations in a wide range of discharge conditions. The maximum relative difference between the number density and mean velocity obtained from the fluid and kinetic simulations is less than 5$\%$ in all cases.
The maximum relative difference for $p$ is less than 10$\%$ in most cases and reaches $\sim 30\%$ only at very low gas pressures
(e.g., at gas pressures below 
1$\,$Pa for the second test problem).\\ \indent
To elucidate further the role of different physical processes we show in Figure~\ref{fig: terms_Ar} the distributions of different terms in the momentum and energy equations for the second test problem at the gas pressures 0.3, 1, 6$\,$Pa and for the benchmark case 1 of Ref.~\cite{turner2013simulation}. The results are presented for the period averaged quantities. As it can be seen from Fig.~\ref{fig: terms_Ar}, one can distinguish between convection- and collision-dominated flows, depending on which terms (convective or collision) compensate the field terms in the momentum and energy equations. Nevertheless, our results demonstrate that it is important to keep the convective, pressure and collision terms together in a wide range of discharge conditions. This allows the model to describe the transition from high to low pressures in a unified way without making assumptions 
about the level of plasma collisionality. Moreover, good quantitative agreement between fluid and PIC-MCC simulations is guaranteed by keeping all terms in the fluid equations. It is also worth noting that the collision-dominated flows in low-pressure discharges can occur in the regime $|u| \gg \upsilon_{0}$. One good example is the results shown in Figs.~\ref{fig: param} and \ref{fig: terms_Ar} for helium at the gas pressure of 4$\,$Pa. It is obvious that an accurate approximation of $\omega$ discussed in Sec.~\ref{Sec: KinModel} is of importance in such cases.
\\ \indent
For completeness, let us also discuss the IDF for the problem under consideration. In Figure~\ref{fig: df_dis} we present the period averaged IDF obtained from the PIC-MCC simulations at different positions in the discharge for the second and third test problems. The results of our PIC-MCC simulations agree qualitatively with those presented in Refs.~\cite{georgieva2004numerical,lee2005ion}. For comparison we also show in Fig.~\ref{fig: df_dis} the IDF given by the model~(\ref{f_band}) and the IDF for drift flows calculated using the model of Ref.~\cite{lampe2012ion} at the local mean velocity. Keeping in mind the results shown in Fig.~\ref{fig: terms_Ar}, one can see that the model~(\ref{f_band}) provides a reasonable qualitative description of the IDF both for the collision- and convection-dominated flows. Furthermore, in the case of 
collision-dominated flows, a more accurate approximation of the IDF can be obtained by employing the model of 
Ref.~\cite{lampe2012ion}.\\ \indent
Considering the results shown in Fig.~\ref{fig: df_dis}, we propose to characterize the ion flow by a simple parameter $\theta=p/p_{d}$, where $p_{d}$ is the internal energy calculated using the IDF for drift flows. According to the discussion of Sec.~\ref{Sec: KinModel}, we get $p_{d}=kT_{a}+(\alpha-1)mu^{2}$, where $\alpha$ is defined in Eq.~(\ref{P}).
Our computations showed that the parameter $\theta$ 
is generally below unity. For $\theta \gtrsim 0.6$
the IDF obtained from the PIC-MCC simulations can be approximated using the model of Ref.~\cite{lampe2012ion}. In this case, the IDF for drift flows deviates noticeably from the simulation results only in the range 
$|\upsilon_{x}| \gg \upsilon_{0}$ (this explains the difference between $p$ and $p_{d}$).
At $\theta \lesssim 0.5$, the ion flow is generally convection-dominated and the IDF for drift flows cannot be used to approximate the IDF for the problem under consideration. As an example, we show in Fig.~\ref{fig: df_dis} the values of $\theta$ calculated for the selected cases.
\\ \indent
\begin{figure}[!t]
\includegraphics[width=0.44\textwidth]{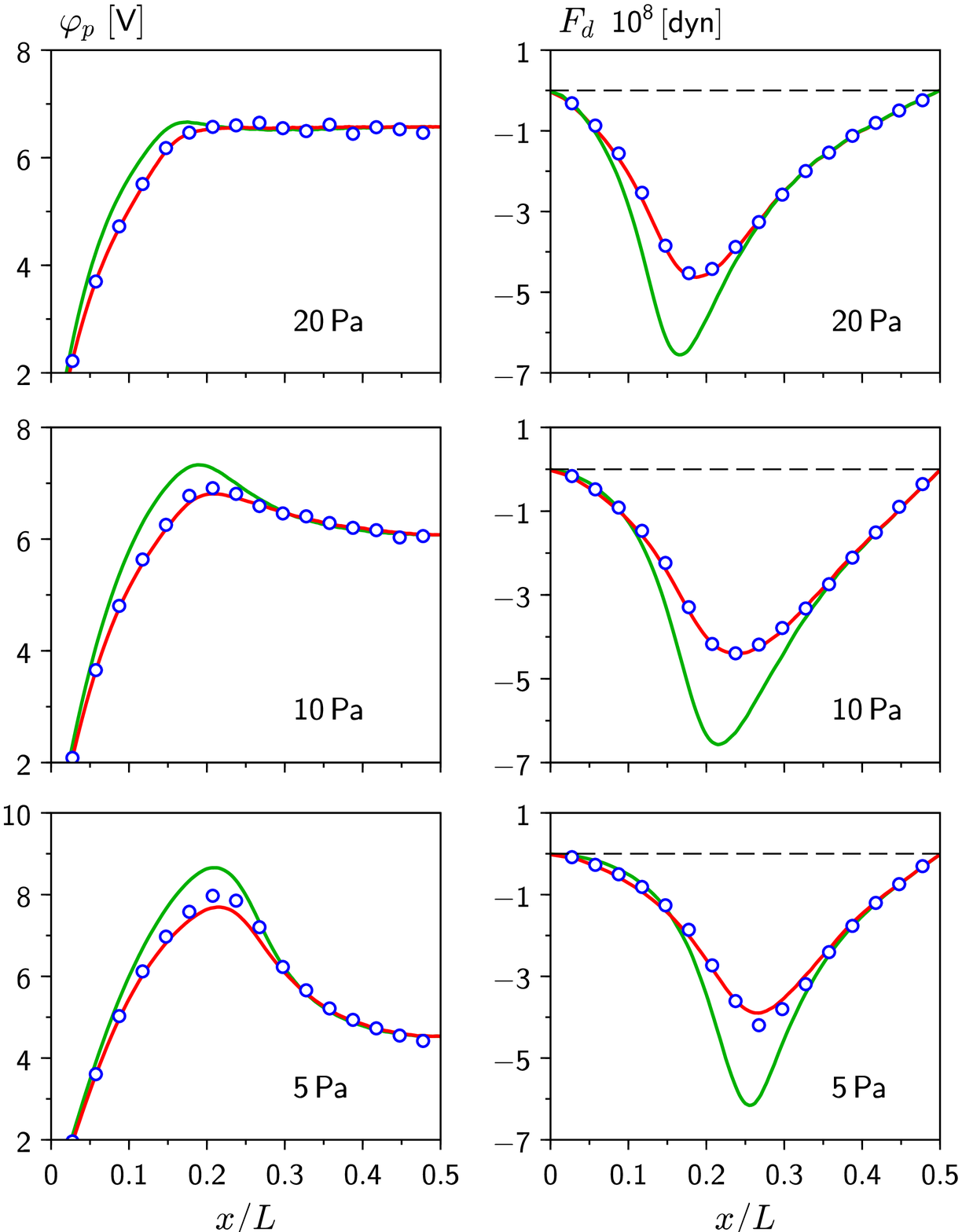}
\caption{\label{fig: dust_Ar} Distributions of the surface potential ($\varphi_{p}$) and the ion drag force ($F_{d}$) calculated for a single dust particle of radius 3.4$\,\mu$m at different positions in the discharge. The gas is argon, the discharge length is 3$\,$cm, the voltage amplitude is $50\,$V and the gas pressures are 20, 10 and 5$\,$Pa (from up to down, respectively). The results were obtained using different models of the IDF. Circles correspond to the IDF taken from the PIC-MCC simulations, red lines correspond to the IDF for drift flows and green lines correspond to the IDF approximated as a shifted Maxwell distribution with the gas temperature.}
\end{figure}
A reasonable approximation of the IDF is important when a more detailed description of the ion component is needed.~One example is modeling of  low-pressure discharges containing dust particles.~Such modeling requires accurate estimate of the ion drag force and rate of ion absorption by the dust. These quantities cannot be reliably calculated without considering the specific form of the IDF. In order to demonstrate this statement, we performed additional computations  for the discharge parameters used in the microgravity experiments with complex plasmas \cite{thomas2008complex} ($L=3\,$cm, the voltage amplitude is 50$\,$V, the gas is argon). In particular, attention was paid to the calculation of the surface potential and the ion drag force for a single dust particle placed at different positions in the discharge. The surface potential was calculated by solving the charging equation described in details in 
Ref.~\cite{fortov2005complex} (see pp.~12-13). The ion drag force was calculated using the general expression presented in Ref.~\cite{khrapak2002ion}. The momentum transfer cross-section for the ion-dust scattering was evaluated using the results of Ref.~\cite{khrapak2014accurate}. The velocity distribution functions for ions and electrons were defined as follows. The electron distribution function was taken from the PIC-MCC simulations. The IDF was evaluated by considering, respectively, the results obtained from the PIC-MCC simulations, the model of Ref.~\cite{lampe2012ion} and the shifted Maxwell distribution function with the gas temperature.\\ \indent
In Figure~\ref{fig: dust_Ar} we show the distributions of the particle surface potential and the ion drag force obtained for the dust particle of radius $3.4\,\mu$m at different gas pressures. It should be noted that for the problem under consideration, the parameter 
$\theta$ was found to be higher than 0.7. Thus, the IDF for drift flows is expected to
be reasonable approximation in this case. 
As it can be seen from Fig.~\ref{fig: dust_Ar}, the specific form of the IDF is of importance. For example, the values of the ion drag force calculated using the shifted Maxwellian distribution deviate substantially from those  obtained using the results of PIC-MCC simulations. On the other hand, the model of Ref.~\cite{lampe2012ion}
gives good agreement with the results of kinetic simulations both for the surface potential and the ion drag force.
\section{Conclusions}
\label{Sec: Concl}
In conclusion, we have proposed a new one-dimensional fluid model for ions in weakly-ionized plasma. Our fluid model differs from previous ones in two aspects. First, we suggest a more accurate approximation of the collision terms in the fluid equations. For this purpose, we use the results obtained  from the MC kinetic simulations of the ion swarm experiments. The proposed approximation, in contrast to the existing ones, can be applied in a wide range of the ion velocities.
Second, we consider the ion energy equation which is closed using a simple model of the IDF.
The accuracy of the fluid equations is examined by comparing with the results of PIC-MCC simulations. In particular, a number of test problems are considered for a parallel plate model of the CCRF discharge. It is shown, that our fluid model gives good agreement with the results of PIC-MCC simulations over a wide range of discharge conditions. In addition, it is shown that the model of the IDF employed in our work can be used to estimate the velocity range where the IDF obtained from the kinetic simulations is essentially non-zero. It is also demonstrated that under certain conditions the IDF obtained from the simulations can be well approximated using the IDF for drift flows evaluated at the local ion mean velocity.
\begin{acknowledgments}
The author is very grateful to Dr.$\,$M.$\,$Yu.$\,$Pustylnik for fruitful discussions and support during the course of this work. 
\end{acknowledgments}
\appendix*
\section{Numerical scheme}
Let us describe the numerical scheme used for solving the governing equations~(\ref{gd_eq}).~First, it can be demonstrated that the flux vector satisfies the homogeneity property $\mathbf{F}=\mathrm{A} \mathbf{U}$, where 
$\mathrm{A}=\partial \mathbf{F} / \partial \mathbf{U}$ is the Jacobian matrix.~At $\gamma=3$  the Jacobian matrix
has three real eigenvalues:
$\lambda_{1}=u$, $\lambda_{2,3}=u \pm v_{c}$. Moreover, $\mathrm{A}$ can be expressed as
$\mathrm{A}=\mathrm {K}\Lambda \mathrm{K}^{-1}$
where 
\begin{align*}
\Lambda_{ij}=\delta_{ij}\lambda_{j}, \quad
\mathrm{K}_{ij}=\lambda_{j}^{i-1}, \quad
i,j=1,2,3.
\end{align*}
The flux vector is then expressed as 
$\mathbf{F}=\mathbf{F}^{+}+\mathbf{F}^{-}$,
where $\mathbf{F}^{\pm}=\mathrm{A}^{\pm} \mathbf{U}$,
$\mathrm{A}^{\pm}=\mathrm{K}\Lambda^{\pm} \mathrm{K}^{-1}$ and
$\Lambda^{\pm}_{ij}=\delta_{ij} \left( \lambda_{j} \pm 
|\lambda_{j}| \right)/2$.
Finally we rewrite the governing equations as
\begin{align*}
\label{gd_num}
\frac{\partial \mathbf{U}}{\partial t}+
\frac{\partial \mathbf{F^{+}}}{\partial x}+
\frac{\partial \mathbf{F^{-}}}{\partial x}
=\mathbf{H}.
\end{align*}
These equations can be approximated numerically using various finite-difference schemes \cite{steger1981flux}. The simplest one is the explicit scheme of the first order (both in time and space). If a uniform grid along the $x$-axis is used, this scheme is written as
\begin{align*}
&\mathbf{U}_{l}^{k+1}-\mathbf{U}_{l}^{k}+
(\tau /h)\,\delta_{l} (\mathbf{F}^{+})^{k}
 \\ 
&+(\tau / h)\, \delta_{l} (\mathbf{F}^{-})^{k}
=\tau \, \mathbf{H}_{l}^{k},
\end{align*} 
where $l$ denotes the grid points, $k$ denotes the time moments, $\tau$ is the time step, $h$ is the grid step and
the finite-differences are 
\begin{align*}
&\delta_{l} (\mathbf{F}^{+})^{k}=
(\mathbf{F}^{+})_{l}^{k}-(\mathbf{F}^{+})_{l-1}^{k},
\\
&\delta_{l} (\mathbf{F}^{-})^{k}=
(\mathbf{F}^{-})_{l+1}^{k}-(\mathbf{F}^{-})_{l}^{k}.
\end{align*}
In the interior points of the computational domain more accurate approximations to the derivatives can be used. For example, in the present work we have employed the following second-order approximations:
\begin{align*}
&\delta_{l} (\mathbf{F}^{+})^{k}=
\left [ 3(\mathbf{F}^{+})_{l}^{k}-4(\mathbf{F}^{+})_{l-1}^{k}
+(\mathbf{F}^{+})_{l-2}^{k} \right]/2,
\\
&\delta_{l} (\mathbf{F}^{-})^{k}=\left[
-(\mathbf{F}^{-})_{l+2}^{k}+4(\mathbf{F}^{-})_{l+1}^{k}-3(\mathbf{F}^{-})_{l}^{k} \right]/2.
\end{align*}
The flux-vector splitting scheme described above takes into account the local characteristic solution of the hyperbolic system~(\ref{gd_eq}). The same analysis has to be done for the boundary conditions. For example, a fully absorbing boundary condition can be simply modeled by neglecting the derivative of the flux
for the incoming waves. That is, we assume $\delta_{l}(\mathbf{F}^{+})^{k}=0$, when the ions flow to the absorbing surface with $u<0$, and
$\delta_{l}(\mathbf{F}^{-})^{k}=0$ in the opposite case.

\begin{thebibliography}{53}%
\makeatletter
\providecommand \@ifxundefined [1]{%
 \@ifx{#1\undefined}
}%
\providecommand \@ifnum [1]{%
 \ifnum #1\expandafter \@firstoftwo
 \else \expandafter \@secondoftwo
 \fi
}%
\providecommand \@ifx [1]{%
 \ifx #1\expandafter \@firstoftwo
 \else \expandafter \@secondoftwo
 \fi
}%
\providecommand \natexlab [1]{#1}%
\providecommand \enquote  [1]{``#1''}%
\providecommand \bibnamefont  [1]{#1}%
\providecommand \bibfnamefont [1]{#1}%
\providecommand \citenamefont [1]{#1}%
\providecommand \href@noop [0]{\@secondoftwo}%
\providecommand \href [0]{\begingroup \@sanitize@url \@href}%
\providecommand \@href[1]{\@@startlink{#1}\@@href}%
\providecommand \@@href[1]{\endgroup#1\@@endlink}%
\providecommand \@sanitize@url [0]{\catcode `\\12\catcode `\$12\catcode
  `\&12\catcode `\#12\catcode `\^12\catcode `\_12\catcode `\%12\relax}%
\providecommand \@@startlink[1]{}%
\providecommand \@@endlink[0]{}%
\providecommand \url  [0]{\begingroup\@sanitize@url \@url }%
\providecommand \@url [1]{\endgroup\@href {#1}{\urlprefix }}%
\providecommand \urlprefix  [0]{URL }%
\providecommand \Eprint [0]{\href }%
\providecommand \doibase [0]{http://dx.doi.org/}%
\providecommand \selectlanguage [0]{\@gobble}%
\providecommand \bibinfo  [0]{\@secondoftwo}%
\providecommand \bibfield  [0]{\@secondoftwo}%
\providecommand \translation [1]{[#1]}%
\providecommand \BibitemOpen [0]{}%
\providecommand \bibitemStop [0]{}%
\providecommand \bibitemNoStop [0]{.\EOS\space}%
\providecommand \EOS [0]{\spacefactor3000\relax}%
\providecommand \BibitemShut  [1]{\csname bibitem#1\endcsname}%
\let\auto@bib@innerbib\@empty
\bibitem [{\citenamefont {Golant}\ \emph {et~al.}(1980)\citenamefont {Golant},
  \citenamefont {Zhilinskii},\ and\ \citenamefont
  {Sakharov}}]{golant1980fundamentals}%
  \BibitemOpen
  \bibfield  {author} {\bibinfo {author} {\bibfnamefont {V.~E.}\ \bibnamefont
  {Golant}}, \bibinfo {author} {\bibfnamefont {A.~P.}\ \bibnamefont
  {Zhilinskii}}, \ and\ \bibinfo {author} {\bibfnamefont {I.~E.}\ \bibnamefont
  {Sakharov}},\ }\href@noop {} {\emph {\bibinfo {title} {Fundamentals of plasma
  physics}}}\ (\bibinfo  {publisher} {Wiley, New York},\ \bibinfo {year}
  {1980})\BibitemShut {NoStop}%
\bibitem [{\citenamefont {Zhdanov}(2002)}]{zhdanov2002transport}%
  \BibitemOpen
  \bibfield  {author} {\bibinfo {author} {\bibfnamefont {V.~M.}\ \bibnamefont
  {Zhdanov}},\ }\href@noop {} {\emph {\bibinfo {title} {Transport Processes in
  Multicomponent Plasma}}}\ (\bibinfo  {publisher} {CRC Press},\ \bibinfo
  {year} {2002})\BibitemShut {NoStop}%
\bibitem [{\citenamefont {Kim}\ \emph {et~al.}(2005{\natexlab{a}})\citenamefont
  {Kim}, \citenamefont {Iza}, \citenamefont {Yang}, \citenamefont
  {Radmilovi{\'c}-Radjenovi{\'c}},\ and\ \citenamefont
  {Lee}}]{kim2005particle}%
  \BibitemOpen
  \bibfield  {author} {\bibinfo {author} {\bibfnamefont {H.~C.}\ \bibnamefont
  {Kim}}, \bibinfo {author} {\bibfnamefont {F.}~\bibnamefont {Iza}}, \bibinfo
  {author} {\bibfnamefont {S.~S.}\ \bibnamefont {Yang}}, \bibinfo {author}
  {\bibfnamefont {M.}~\bibnamefont {Radmilovi{\'c}-Radjenovi{\'c}}}, \ and\
  \bibinfo {author} {\bibfnamefont {J.~K.}\ \bibnamefont {Lee}},\ }\href@noop
  {} {\bibfield  {journal} {\bibinfo  {journal} {J. Phys. D: Appl. Phys.}\
  }\textbf {\bibinfo {volume} {38}},\ \bibinfo {pages} {R283} (\bibinfo {year}
  {2005}{\natexlab{a}})}\BibitemShut {NoStop}%
\bibitem [{\citenamefont {Tsendin}(1995)}]{tsendin1995electron}%
  \BibitemOpen
  \bibfield  {author} {\bibinfo {author} {\bibfnamefont {L.~D.}\ \bibnamefont
  {Tsendin}},\ }\href@noop {} {\bibfield  {journal} {\bibinfo  {journal}
  {Plasma Sources Sci. Technol.}\ }\textbf {\bibinfo {volume} {4}},\ \bibinfo
  {pages} {200} (\bibinfo {year} {1995})}\BibitemShut {NoStop}%
\bibitem [{\citenamefont {Hagelaar}\ and\ \citenamefont
  {Pitchford}(2005)}]{hagelaar2005solving}%
  \BibitemOpen
  \bibfield  {author} {\bibinfo {author} {\bibfnamefont {G.~J.~M.}\
  \bibnamefont {Hagelaar}}\ and\ \bibinfo {author} {\bibfnamefont {L.~C.}\
  \bibnamefont {Pitchford}},\ }\href@noop {} {\bibfield  {journal} {\bibinfo
  {journal} {Plasma Sources Sci. Technol.}\ }\textbf {\bibinfo {volume} {14}},\
  \bibinfo {pages} {722} (\bibinfo {year} {2005})}\BibitemShut {NoStop}%
\bibitem [{\citenamefont {Kolobov}\ and\ \citenamefont
  {Arslanbekov}(2006)}]{kolobov2006simulation}%
  \BibitemOpen
  \bibfield  {author} {\bibinfo {author} {\bibfnamefont {V.~I.}\ \bibnamefont
  {Kolobov}}\ and\ \bibinfo {author} {\bibfnamefont {R.~R.}\ \bibnamefont
  {Arslanbekov}},\ }\href@noop {} {\bibfield  {journal} {\bibinfo  {journal}
  {IEEE Trans. Plasma Sci.}\ }\textbf {\bibinfo {volume} {34}},\ \bibinfo
  {pages} {895} (\bibinfo {year} {2006})}\BibitemShut {NoStop}%
\bibitem [{\citenamefont {Richards}\ \emph {et~al.}(1987)\citenamefont
  {Richards}, \citenamefont {Thompson},\ and\ \citenamefont
  {Sawin}}]{richards1987continuum}%
  \BibitemOpen
  \bibfield  {author} {\bibinfo {author} {\bibfnamefont {A.~D.}\ \bibnamefont
  {Richards}}, \bibinfo {author} {\bibfnamefont {B.~E.}\ \bibnamefont
  {Thompson}}, \ and\ \bibinfo {author} {\bibfnamefont {H.~H.}\ \bibnamefont
  {Sawin}},\ }\href@noop {} {\bibfield  {journal} {\bibinfo  {journal} {Appl.
  Phys. Lett.}\ }\textbf {\bibinfo {volume} {50}},\ \bibinfo {pages} {492}
  (\bibinfo {year} {1987})}\BibitemShut {NoStop}%
\bibitem [{\citenamefont {Meyyappan}\ and\ \citenamefont
  {Kreskovsky}(1990)}]{meyyappan1990glow}%
  \BibitemOpen
  \bibfield  {author} {\bibinfo {author} {\bibfnamefont {M.}~\bibnamefont
  {Meyyappan}}\ and\ \bibinfo {author} {\bibfnamefont {J.~P.}\ \bibnamefont
  {Kreskovsky}},\ }\href@noop {} {\bibfield  {journal} {\bibinfo  {journal} {J.
  Appl. Phys.}\ }\textbf {\bibinfo {volume} {68}},\ \bibinfo {pages} {1506}
  (\bibinfo {year} {1990})}\BibitemShut {NoStop}%
\bibitem [{\citenamefont {Gogolides}\ and\ \citenamefont
  {Sawin}(1992{\natexlab{a}})}]{gogolides1992continuum1}%
  \BibitemOpen
  \bibfield  {author} {\bibinfo {author} {\bibfnamefont {E.}~\bibnamefont
  {Gogolides}}\ and\ \bibinfo {author} {\bibfnamefont {H.~H.}\ \bibnamefont
  {Sawin}},\ }\href@noop {} {\bibfield  {journal} {\bibinfo  {journal} {J.
  Appl. Phys.}\ }\textbf {\bibinfo {volume} {72}},\ \bibinfo {pages} {3971}
  (\bibinfo {year} {1992}{\natexlab{a}})}\BibitemShut {NoStop}%
\bibitem [{\citenamefont {Gogolides}\ and\ \citenamefont
  {Sawin}(1992{\natexlab{b}})}]{gogolides1992continuum2}%
  \BibitemOpen
  \bibfield  {author} {\bibinfo {author} {\bibfnamefont {E.}~\bibnamefont
  {Gogolides}}\ and\ \bibinfo {author} {\bibfnamefont {H.~H.}\ \bibnamefont
  {Sawin}},\ }\href@noop {} {\bibfield  {journal} {\bibinfo  {journal} {J.
  Appl. Phys.}\ }\textbf {\bibinfo {volume} {72}},\ \bibinfo {pages} {3988}
  (\bibinfo {year} {1992}{\natexlab{b}})}\BibitemShut {NoStop}%
\bibitem [{\citenamefont {Passchier}\ and\ \citenamefont
  {Goedheer}(1993)}]{passchier1993two}%
  \BibitemOpen
  \bibfield  {author} {\bibinfo {author} {\bibfnamefont {J.~D.~P.}\
  \bibnamefont {Passchier}}\ and\ \bibinfo {author} {\bibfnamefont {W.~J.}\
  \bibnamefont {Goedheer}},\ }\href@noop {} {\bibfield  {journal} {\bibinfo
  {journal} {J. Appl. Phys.}\ }\textbf {\bibinfo {volume} {74}},\ \bibinfo
  {pages} {3744} (\bibinfo {year} {1993})}\BibitemShut {NoStop}%
\bibitem [{\citenamefont {Nitschke}\ and\ \citenamefont
  {Graves}(1994)}]{nitschke1994comparison}%
  \BibitemOpen
  \bibfield  {author} {\bibinfo {author} {\bibfnamefont {T.~E.}\ \bibnamefont
  {Nitschke}}\ and\ \bibinfo {author} {\bibfnamefont {D.~B.}\ \bibnamefont
  {Graves}},\ }\href@noop {} {\bibfield  {journal} {\bibinfo  {journal} {J.
  Appl. Phys.}\ }\textbf {\bibinfo {volume} {76}},\ \bibinfo {pages} {5646}
  (\bibinfo {year} {1994})}\BibitemShut {NoStop}%
\bibitem [{\citenamefont {Boeuf}\ and\ \citenamefont
  {Pitchford}(1995)}]{boeuf1995two}%
  \BibitemOpen
  \bibfield  {author} {\bibinfo {author} {\bibfnamefont {J.~P.}\ \bibnamefont
  {Boeuf}}\ and\ \bibinfo {author} {\bibfnamefont {L.~C.}\ \bibnamefont
  {Pitchford}},\ }\href@noop {} {\bibfield  {journal} {\bibinfo  {journal}
  {Phys. Rev. E}\ }\textbf {\bibinfo {volume} {51}},\ \bibinfo {pages} {1376}
  (\bibinfo {year} {1995})}\BibitemShut {NoStop}%
\bibitem [{\citenamefont {Becker}\ \emph {et~al.}(2016)\citenamefont {Becker},
  \citenamefont {K{\"a}hlert}, \citenamefont {Sun}, \citenamefont {Bonitz},\
  and\ \citenamefont {Loffhagen}}]{becker2016advanced}%
  \BibitemOpen
  \bibfield  {author} {\bibinfo {author} {\bibfnamefont {M.~M.}\ \bibnamefont
  {Becker}}, \bibinfo {author} {\bibfnamefont {H.}~\bibnamefont {K{\"a}hlert}},
  \bibinfo {author} {\bibfnamefont {A.}~\bibnamefont {Sun}}, \bibinfo {author}
  {\bibfnamefont {M.}~\bibnamefont {Bonitz}}, \ and\ \bibinfo {author}
  {\bibfnamefont {D.}~\bibnamefont {Loffhagen}},\ }\href@noop {} {\bibfield
  {journal} {\bibinfo  {journal} {arXiv preprint arXiv:1608.04601}\ } (\bibinfo
  {year} {2016})}\BibitemShut {NoStop}%
\bibitem [{\citenamefont {Chen}\ \emph {et~al.}(2016)\citenamefont {Chen},
  \citenamefont {Tseng}, \citenamefont {Gu}, \citenamefont {Hung},\ and\
  \citenamefont {Wu}}]{chen2016numerical}%
  \BibitemOpen
  \bibfield  {author} {\bibinfo {author} {\bibfnamefont {K.~L.}\ \bibnamefont
  {Chen}}, \bibinfo {author} {\bibfnamefont {M.~F.}\ \bibnamefont {Tseng}},
  \bibinfo {author} {\bibfnamefont {B.~R.}\ \bibnamefont {Gu}}, \bibinfo
  {author} {\bibfnamefont {C.~T.}\ \bibnamefont {Hung}}, \ and\ \bibinfo
  {author} {\bibfnamefont {J.~S.}\ \bibnamefont {Wu}},\ }\href@noop {}
  {\bibfield  {journal} {\bibinfo  {journal} {IEEE Trans. Plasma Sci.}\ }\textbf {\bibinfo {volume} {PP}},\ \bibinfo {pages} {1-8}
  (\bibinfo {year} {2016})}\BibitemShut {NoStop}%
\bibitem [{\citenamefont {Godyak}(1982)}]{godyak1982modified}%
  \BibitemOpen
  \bibfield  {author} {\bibinfo {author} {\bibfnamefont {V.~A.}\ \bibnamefont
  {Godyak}},\ }\href@noop {} {\bibfield  {journal} {\bibinfo  {journal} {Phys.
  Lett. A}\ }\textbf {\bibinfo {volume} {89}},\ \bibinfo {pages} {80} (\bibinfo
  {year} {1982})}\BibitemShut {NoStop}%
\bibitem [{\citenamefont {Sternberg}\ and\ \citenamefont
  {Godyak}(2003)}]{sternberg2003patching}%
  \BibitemOpen
  \bibfield  {author} {\bibinfo {author} {\bibfnamefont {N.}~\bibnamefont
  {Sternberg}}\ and\ \bibinfo {author} {\bibfnamefont {V.}~\bibnamefont
  {Godyak}},\ }\href@noop {} {\bibfield  {journal} {\bibinfo  {journal} {IEEE
  Trans. Plasma Sci.}\ }\textbf {\bibinfo {volume} {31}},\ \bibinfo {pages}
  {1395} (\bibinfo {year} {2003})}\BibitemShut {NoStop}%
\bibitem [{\citenamefont {Sternberg}\ and\ \citenamefont
  {Godyak}(2007)}]{sternberg2007bohm}%
  \BibitemOpen
  \bibfield  {author} {\bibinfo {author} {\bibfnamefont {N.}~\bibnamefont
  {Sternberg}}\ and\ \bibinfo {author} {\bibfnamefont {V.}~\bibnamefont
  {Godyak}},\ }\href@noop {} {\bibfield  {journal} {\bibinfo  {journal} {IEEE
  Trans. Plasma Sci.}\ }\textbf {\bibinfo {volume} {35}},\ \bibinfo {pages}
  {1341} (\bibinfo {year} {2007})}\BibitemShut {NoStop}%
\bibitem [{\citenamefont {Brinkmann}(2011)}]{brinkmann2011plasma}%
  \BibitemOpen
  \bibfield  {author} {\bibinfo {author} {\bibfnamefont {R.~P.}\ \bibnamefont
  {Brinkmann}},\ }\href@noop {} {\bibfield  {journal} {\bibinfo  {journal} {J.
  Phys. D: Appl. Phys.}\ }\textbf {\bibinfo {volume} {44}},\ \bibinfo {pages}
  {042002} (\bibinfo {year} {2011})}\BibitemShut {NoStop}%
\bibitem [{\citenamefont {Surendra}\ and\ \citenamefont
  {Dalvie}(1993)}]{surendra1993moment}%
  \BibitemOpen
  \bibfield  {author} {\bibinfo {author} {\bibfnamefont {M.}~\bibnamefont
  {Surendra}}\ and\ \bibinfo {author} {\bibfnamefont {M.}~\bibnamefont
  {Dalvie}},\ }\href@noop {} {\bibfield  {journal} {\bibinfo  {journal} {Phys.
  Rev. E}\ }\textbf {\bibinfo {volume} {48}},\ \bibinfo {pages} {3914}
  (\bibinfo {year} {1993})}\BibitemShut {NoStop}%
\bibitem [{\citenamefont {Ellis}\ \emph {et~al.}(1976)\citenamefont {Ellis},
  \citenamefont {Pai}, \citenamefont {McDaniel}, \citenamefont {Mason},\ and\
  \citenamefont {Viehland}}]{ellis1976transport}%
  \BibitemOpen
  \bibfield  {author} {\bibinfo {author} {\bibfnamefont {H.~W.}\ \bibnamefont
  {Ellis}}, \bibinfo {author} {\bibfnamefont {R.~Y.}\ \bibnamefont {Pai}},
  \bibinfo {author} {\bibfnamefont {E.~W.}\ \bibnamefont {McDaniel}}, \bibinfo
  {author} {\bibfnamefont {E.~A.}\ \bibnamefont {Mason}}, \ and\ \bibinfo
  {author} {\bibfnamefont {L.~A.}\ \bibnamefont {Viehland}},\ }\href@noop {}
  {\bibfield  {journal} {\bibinfo  {journal} {At. Data Nucl. Data Tables}\
  }\textbf {\bibinfo {volume} {17}},\ \bibinfo {pages} {177} (\bibinfo {year}
  {1976})}\BibitemShut {NoStop}%
\bibitem [{\citenamefont {Phelps}(1991)}]{phelps1991cross}%
  \BibitemOpen
  \bibfield  {author} {\bibinfo {author} {\bibfnamefont {A.~V.}\ \bibnamefont
  {Phelps}},\ }\href@noop {} {\bibfield  {journal} {\bibinfo  {journal} {J.
  Phys. Chem. Ref. Data}\ }\textbf {\bibinfo {volume} {20}},\ \bibinfo {pages}
  {557} (\bibinfo {year} {1991})}\BibitemShut {NoStop}%
\bibitem [{\citenamefont {Vahedi}\ and\ \citenamefont
  {Surendra}(1995)}]{vahedi1995monte}%
  \BibitemOpen
  \bibfield  {author} {\bibinfo {author} {\bibfnamefont {V.}~\bibnamefont
  {Vahedi}}\ and\ \bibinfo {author} {\bibfnamefont {M.}~\bibnamefont
  {Surendra}},\ }\href@noop {} {\bibfield  {journal} {\bibinfo  {journal}
  {Comput. Phys. Commun.}\ }\textbf {\bibinfo {volume} {87}},\ \bibinfo {pages}
  {179} (\bibinfo {year} {1995})}\BibitemShut {NoStop}%
\bibitem [{\citenamefont {Verboncoeur}(2005)}]{verboncoeur2005particle}%
  \BibitemOpen
  \bibfield  {author} {\bibinfo {author} {\bibfnamefont {J.~P.}\ \bibnamefont
  {Verboncoeur}},\ }\href@noop {} {\bibfield  {journal} {\bibinfo  {journal}
  {Plasma Phys. Control. Fusion}\ }\textbf {\bibinfo {volume} {47}},\ \bibinfo
  {pages} {A231} (\bibinfo {year} {2005})}\BibitemShut {NoStop}%
\bibitem [{\citenamefont {Donk{\'o}}(2011)}]{donko2011particle}%
  \BibitemOpen
  \bibfield  {author} {\bibinfo {author} {\bibfnamefont {Z.}~\bibnamefont
  {Donk{\'o}}},\ }\href@noop {} {\bibfield  {journal} {\bibinfo  {journal}
  {Plasma Sources Sci. Technol.}\ }\textbf {\bibinfo {volume} {20}},\ \bibinfo
  {pages} {024001} (\bibinfo {year} {2011})}\BibitemShut {NoStop}%
\bibitem [{\citenamefont {Turner}\ \emph {et~al.}(2013)\citenamefont {Turner},
  \citenamefont {Derzsi}, \citenamefont {Donko}, \citenamefont {Eremin},
  \citenamefont {Kelly}, \citenamefont {Lafleur},\ and\ \citenamefont
  {Mussenbrock}}]{turner2013simulation}%
  \BibitemOpen
  \bibfield  {author} {\bibinfo {author} {\bibfnamefont {M.~M.}\ \bibnamefont
  {Turner}}, \bibinfo {author} {\bibfnamefont {A.}~\bibnamefont {Derzsi}},
  \bibinfo {author} {\bibfnamefont {Z.}~\bibnamefont {Donko}}, \bibinfo
  {author} {\bibfnamefont {D.}~\bibnamefont {Eremin}}, \bibinfo {author}
  {\bibfnamefont {S.~J.}\ \bibnamefont {Kelly}}, \bibinfo {author}
  {\bibfnamefont {T.}~\bibnamefont {Lafleur}}, \ and\ \bibinfo {author}
  {\bibfnamefont {T.}~\bibnamefont {Mussenbrock}},\ }\href@noop {} {\bibfield
  {journal} {\bibinfo  {journal} {Phys. Plasmas}\ }\textbf {\bibinfo {volume}
  {20}},\ \bibinfo {pages} {013507} (\bibinfo {year} {2013})}\BibitemShut
  {NoStop}%
\bibitem [{\citenamefont {Godyak}\ and\ \citenamefont
  {Piejak}(1990)}]{godyak1990abnormally}%
  \BibitemOpen
  \bibfield  {author} {\bibinfo {author} {\bibfnamefont {V.~A.}\ \bibnamefont
  {Godyak}}\ and\ \bibinfo {author} {\bibfnamefont {R.~B.}\ \bibnamefont
  {Piejak}},\ }\href@noop {} {\bibfield  {journal} {\bibinfo  {journal} {Phys.
  Rev. Lett.}\ }\textbf {\bibinfo {volume} {65}},\ \bibinfo {pages} {996}
  (\bibinfo {year} {1990})}\BibitemShut {NoStop}%
\bibitem [{\citenamefont {Godyak}\ \emph {et~al.}(1992)\citenamefont {Godyak},
  \citenamefont {Piejak},\ and\ \citenamefont
  {Alexandrovich}}]{godyak1992measurement}%
  \BibitemOpen
  \bibfield  {author} {\bibinfo {author} {\bibfnamefont {V.~A.}\ \bibnamefont
  {Godyak}}, \bibinfo {author} {\bibfnamefont {R.~B.}\ \bibnamefont {Piejak}},
  \ and\ \bibinfo {author} {\bibfnamefont {B.~M.}\ \bibnamefont
  {Alexandrovich}},\ }\href@noop {} {\bibfield  {journal} {\bibinfo  {journal}
  {Plasma Sources Sci. Technol.}\ }\textbf {\bibinfo {volume} {1}},\ \bibinfo
  {pages} {36} (\bibinfo {year} {1992})}\BibitemShut {NoStop}%
\bibitem [{\citenamefont {Lampe}\ \emph {et~al.}(2012)\citenamefont {Lampe},
  \citenamefont {R{\"o}cker}, \citenamefont {Joyce}, \citenamefont {Zhdanov},
  \citenamefont {Ivlev},\ and\ \citenamefont {Morfill}}]{lampe2012ion}%
  \BibitemOpen
  \bibfield  {author} {\bibinfo {author} {\bibfnamefont {M.}~\bibnamefont
  {Lampe}}, \bibinfo {author} {\bibfnamefont {T.~B.}\ \bibnamefont
  {R{\"o}cker}}, \bibinfo {author} {\bibfnamefont {G.}~\bibnamefont {Joyce}},
  \bibinfo {author} {\bibfnamefont {S.~K.}\ \bibnamefont {Zhdanov}}, \bibinfo
  {author} {\bibfnamefont {A.~V.}\ \bibnamefont {Ivlev}}, \ and\ \bibinfo
  {author} {\bibfnamefont {G.~E.}\ \bibnamefont {Morfill}},\ }\href@noop {}
  {\bibfield  {journal} {\bibinfo  {journal} {Phys. Plasmas}\ }\textbf
  {\bibinfo {volume} {19}},\ \bibinfo {pages} {113703} (\bibinfo {year}
  {2012})}\BibitemShut {NoStop}%
\bibitem [{LXC()}]{LXCat}%
  \BibitemOpen
  \href@noop {} {}\bibinfo {howpublished}
  {\url{www.lxcat.laplace.univ-tlse.fr}}\BibitemShut {NoStop}%
\bibitem [{\citenamefont {Phelps}(1994)}]{phelps1994application}%
  \BibitemOpen
  \bibfield  {author} {\bibinfo {author} {\bibfnamefont {A.~V.}\ \bibnamefont
  {Phelps}},\ }\href@noop {} {\bibfield  {journal} {\bibinfo  {journal} {J.
  Appl. Phys.}\ }\textbf {\bibinfo {volume} {76}},\ \bibinfo {pages} {747}
  (\bibinfo {year} {1994})}\BibitemShut {NoStop}%
\bibitem [{\citenamefont {Devoto}(1967)}]{devoto1967transport}%
  \BibitemOpen
  \bibfield  {author} {\bibinfo {author} {\bibfnamefont {R.~S.}\ \bibnamefont
  {Devoto}},\ }\href@noop {} {\bibfield  {journal} {\bibinfo  {journal} {Phys.
  Fluids}\ }\textbf {\bibinfo {volume} {10}},\ \bibinfo {pages} {354} (\bibinfo
  {year} {1967})}\BibitemShut {NoStop}%
\bibitem [{\citenamefont {Ziegler}(1953)}]{ziegler1953wirkungsquerschnitt}%
  \BibitemOpen
  \bibfield  {author} {\bibinfo {author} {\bibfnamefont {B.}~\bibnamefont
  {Ziegler}},\ }\href@noop {} {\bibfield  {journal} {\bibinfo  {journal} {Z.
  Physik}\ }\textbf {\bibinfo {volume} {136}},\ \bibinfo {pages} {108}
  (\bibinfo {year} {1953})}\BibitemShut {NoStop}%
\bibitem [{\citenamefont {Cramer}(1959)}]{cramer1959elastic}%
  \BibitemOpen
  \bibfield  {author} {\bibinfo {author} {\bibfnamefont {W.~H.}\ \bibnamefont
  {Cramer}},\ }\href@noop {} {\bibfield  {journal} {\bibinfo  {journal} {J.
  Chem. Phys.}\ }\textbf {\bibinfo {volume} {30}},\ \bibinfo {pages} {641}
  (\bibinfo {year} {1959})}\BibitemShut {NoStop}%
\bibitem [{\citenamefont {Khrapak}(2013)}]{khrapak2013practical}%
  \BibitemOpen
  \bibfield  {author} {\bibinfo {author} {\bibfnamefont {S.~A.}\ \bibnamefont
  {Khrapak}},\ }\href@noop {} {\bibfield  {journal} {\bibinfo  {journal} {J.
  Plasma Physics}\ }\textbf {\bibinfo {volume} {79}},\ \bibinfo {pages} {1123}
  (\bibinfo {year} {2013})}\BibitemShut {NoStop}%
\bibitem [{\citenamefont {Frost}(1957)}]{frost1957effect}%
  \BibitemOpen
  \bibfield  {author} {\bibinfo {author} {\bibfnamefont {L.~S.}\ \bibnamefont
  {Frost}},\ }\href@noop {} {\bibfield  {journal} {\bibinfo  {journal} {Phys.
  Rev.}\ }\textbf {\bibinfo {volume} {105}},\ \bibinfo {pages} {354} (\bibinfo
  {year} {1957})}\BibitemShut {NoStop}%
\bibitem [{\citenamefont {Lawler}(1985)}]{lawler1985equilibration}%
  \BibitemOpen
  \bibfield  {author} {\bibinfo {author} {\bibfnamefont {J.~E.}\ \bibnamefont
  {Lawler}},\ }\href@noop {} {\bibfield  {journal} {\bibinfo  {journal} {Phys.
  Rev. A}\ }\textbf {\bibinfo {volume} {32}},\ \bibinfo {pages} {2977}
  (\bibinfo {year} {1985})}\BibitemShut {NoStop}%
\bibitem [{\citenamefont {Stangeby}(1984)}]{stangeby1984plasma}%
  \BibitemOpen
  \bibfield  {author} {\bibinfo {author} {\bibfnamefont {P.~C.}\ \bibnamefont
  {Stangeby}},\ }\href@noop {} {\bibfield  {journal} {\bibinfo  {journal}
  {Phys. Fluids}\ }\textbf {\bibinfo {volume} {27}},\ \bibinfo {pages} {682}
  (\bibinfo {year} {1984})}\BibitemShut {NoStop}%
\bibitem [{\citenamefont {Benilov}\ and\ \citenamefont
  {Marotta}(1995)}]{benilov1995model}%
  \BibitemOpen
  \bibfield  {author} {\bibinfo {author} {\bibfnamefont {M.~S.}\ \bibnamefont
  {Benilov}}\ and\ \bibinfo {author} {\bibfnamefont {A.}~\bibnamefont
  {Marotta}},\ }\href@noop {} {\bibfield  {journal} {\bibinfo  {journal} {J.
  Phys. D: Appl. Phys.}\ }\textbf {\bibinfo {volume} {28}},\ \bibinfo {pages}
  {1869} (\bibinfo {year} {1995})}\BibitemShut {NoStop}%
\bibitem [{\citenamefont {LeVeque}(2002)}]{leveque2002finite}%
  \BibitemOpen
  \bibfield  {author} {\bibinfo {author} {\bibfnamefont {R.~J.}\ \bibnamefont
  {LeVeque}},\ }\href@noop {} {\emph {\bibinfo {title} {Finite volume methods
  for hyperbolic problems}}}\ (\bibinfo  {publisher} {Cambridge university
  press},\ \bibinfo {year} {2002})\BibitemShut {NoStop}%
\bibitem [{\citenamefont {Toro}(2013)}]{toro2013riemann}%
  \BibitemOpen
  \bibfield  {author} {\bibinfo {author} {\bibfnamefont {E.~F.}\ \bibnamefont
  {Toro}},\ }\href@noop {} {\emph {\bibinfo {title} {Riemann solvers and
  numerical methods for fluid dynamics: a practical introduction}}}\ (\bibinfo
  {publisher} {Springer Science \& Business Media},\ \bibinfo {year}
  {2013})\BibitemShut {NoStop}%
\bibitem [{\citenamefont {Steger}\ and\ \citenamefont
  {Warming}(1981)}]{steger1981flux}%
  \BibitemOpen
  \bibfield  {author} {\bibinfo {author} {\bibfnamefont {J.~L.}\ \bibnamefont
  {Steger}}\ and\ \bibinfo {author} {\bibfnamefont {R.~F.}\ \bibnamefont
  {Warming}},\ }\href@noop {} {\bibfield  {journal} {\bibinfo  {journal} {J.
  Computational Phys.}\ }\textbf {\bibinfo {volume} {40}},\ \bibinfo {pages}
  {263} (\bibinfo {year} {1981})}\BibitemShut {NoStop}%
\bibitem [{\citenamefont {Vahedi}\ \emph
  {et~al.}(1993{\natexlab{a}})\citenamefont {Vahedi}, \citenamefont {DiPeso},
  \citenamefont {Birdsall}, \citenamefont {Lieberman},\ and\ \citenamefont
  {Rognlien}}]{vahedi1993capacitive1}%
  \BibitemOpen
  \bibfield  {author} {\bibinfo {author} {\bibfnamefont {V.}~\bibnamefont
  {Vahedi}}, \bibinfo {author} {\bibfnamefont {G.}~\bibnamefont {DiPeso}},
  \bibinfo {author} {\bibfnamefont {C.~K.}\ \bibnamefont {Birdsall}}, \bibinfo
  {author} {\bibfnamefont {M.~A.}\ \bibnamefont {Lieberman}}, \ and\ \bibinfo
  {author} {\bibfnamefont {T.~D.}\ \bibnamefont {Rognlien}},\ }\href@noop {}
  {\bibfield  {journal} {\bibinfo  {journal} {Plasma Sources Sci. Technol.}\
  }\textbf {\bibinfo {volume} {2}},\ \bibinfo {pages} {261} (\bibinfo {year}
  {1993}{\natexlab{a}})}\BibitemShut {NoStop}%
\bibitem [{\citenamefont {Vahedi}\ \emph
  {et~al.}(1993{\natexlab{b}})\citenamefont {Vahedi}, \citenamefont {DiPeso},
  \citenamefont {Birdsall}, \citenamefont {Lieberman},\ and\ \citenamefont
  {Rognlien}}]{vahedi1993capacitive2}%
  \BibitemOpen
  \bibfield  {author} {\bibinfo {author} {\bibfnamefont {V.}~\bibnamefont
  {Vahedi}}, \bibinfo {author} {\bibfnamefont {G.}~\bibnamefont {DiPeso}},
  \bibinfo {author} {\bibfnamefont {C.~K.}\ \bibnamefont {Birdsall}}, \bibinfo
  {author} {\bibfnamefont {M.~A.}\ \bibnamefont {Lieberman}}, \ and\ \bibinfo
  {author} {\bibfnamefont {T.~D.}\ \bibnamefont {Rognlien}},\ }\href@noop {}
  {\bibfield  {journal} {\bibinfo  {journal} {Plasma Sources Sci. Technol.}\
  }\textbf {\bibinfo {volume} {2}},\ \bibinfo {pages} {273} (\bibinfo {year}
  {1993}{\natexlab{b}})}\BibitemShut {NoStop}%
\bibitem [{\citenamefont {Kim}\ \emph {et~al.}(2005{\natexlab{b}})\citenamefont
  {Kim}, \citenamefont {Manuilenko},\ and\ \citenamefont {Lee}}]{kim2005PIC}%
  \BibitemOpen
  \bibfield  {author} {\bibinfo {author} {\bibfnamefont {H.~C.}\ \bibnamefont
  {Kim}}, \bibinfo {author} {\bibfnamefont {O.}~\bibnamefont {Manuilenko}}, \
  and\ \bibinfo {author} {\bibfnamefont {J.~K.}\ \bibnamefont {Lee}},\
  }\href@noop {} {\bibfield  {journal} {\bibinfo  {journal} {Japan. J. Appl.
  Phys.}\ }\textbf {\bibinfo {volume} {44}},\ \bibinfo {pages} {1957} (\bibinfo
  {year} {2005}{\natexlab{b}})}\BibitemShut {NoStop}%
\bibitem [{\citenamefont {Smith}(1930)}]{smith1930ionization}%
  \BibitemOpen
  \bibfield  {author} {\bibinfo {author} {\bibfnamefont {P.~T.}\ \bibnamefont
  {Smith}},\ }\href@noop {} {\bibfield  {journal} {\bibinfo  {journal} {Phys.
  Rev.}\ }\textbf {\bibinfo {volume} {36}},\ \bibinfo {pages} {1293} (\bibinfo
  {year} {1930})}\BibitemShut {NoStop}%
\bibitem [{\citenamefont {Okhrimovskyy}\ \emph {et~al.}(2002)\citenamefont
  {Okhrimovskyy}, \citenamefont {Bogaerts},\ and\ \citenamefont
  {Gijbels}}]{okhrimovskyy2002electron}%
  \BibitemOpen
  \bibfield  {author} {\bibinfo {author} {\bibfnamefont {A.}~\bibnamefont
  {Okhrimovskyy}}, \bibinfo {author} {\bibfnamefont {A.}~\bibnamefont
  {Bogaerts}}, \ and\ \bibinfo {author} {\bibfnamefont {R.}~\bibnamefont
  {Gijbels}},\ }\href@noop {} {\bibfield  {journal} {\bibinfo  {journal} {Phys.
  Rev. E}\ }\textbf {\bibinfo {volume} {65}},\ \bibinfo {pages} {037402}
  (\bibinfo {year} {2002})}\BibitemShut {NoStop}%
\bibitem [{\citenamefont {Georgieva}\ \emph {et~al.}(2004)\citenamefont
  {Georgieva}, \citenamefont {Bogaerts},\ and\ \citenamefont
  {Gijbels}}]{georgieva2004numerical}%
  \BibitemOpen
  \bibfield  {author} {\bibinfo {author} {\bibfnamefont {V.}~\bibnamefont
  {Georgieva}}, \bibinfo {author} {\bibfnamefont {A.}~\bibnamefont {Bogaerts}},
  \ and\ \bibinfo {author} {\bibfnamefont {R.}~\bibnamefont {Gijbels}},\
  }\href@noop {} {\bibfield  {journal} {\bibinfo  {journal} {Phys. Rev. E}\
  }\textbf {\bibinfo {volume} {69}},\ \bibinfo {pages} {026406} (\bibinfo
  {year} {2004})}\BibitemShut {NoStop}%
\bibitem [{\citenamefont {Lee}\ \emph {et~al.}(2005)\citenamefont {Lee},
  \citenamefont {Manuilenko}, \citenamefont {Babaeva}, \citenamefont {Kim},\
  and\ \citenamefont {Shon}}]{lee2005ion}%
  \BibitemOpen
  \bibfield  {author} {\bibinfo {author} {\bibfnamefont {J.~K.}\ \bibnamefont
  {Lee}}, \bibinfo {author} {\bibfnamefont {O.~V.}\ \bibnamefont {Manuilenko}},
  \bibinfo {author} {\bibfnamefont {N.~Y.}\ \bibnamefont {Babaeva}}, \bibinfo
  {author} {\bibfnamefont {H.~C.}\ \bibnamefont {Kim}}, \ and\ \bibinfo
  {author} {\bibfnamefont {J.~W.}\ \bibnamefont {Shon}},\ }\href@noop {}
  {\bibfield  {journal} {\bibinfo  {journal} {Plasma Sources Sci. Technol.}\
  }\textbf {\bibinfo {volume} {14}},\ \bibinfo {pages} {89} (\bibinfo {year}
  {2005})}\BibitemShut {NoStop}%
\bibitem [{\citenamefont {Thomas}\ \emph {et~al.}(2008)\citenamefont {Thomas},
  \citenamefont {Morfill}, \citenamefont {Fortov} \emph
  {et~al.}}]{thomas2008complex}%
  \BibitemOpen
  \bibfield  {author} {\bibinfo {author} {\bibfnamefont {H.~M.}\ \bibnamefont
  {Thomas}}, \bibinfo {author} {\bibfnamefont {G.~E.}\ \bibnamefont {Morfill}},
  \bibinfo {author} {\bibfnamefont {V.~E.}\ \bibnamefont {Fortov}},  \emph
  {et~al.},\ }\href@noop {} {\bibfield  {journal} {\bibinfo  {journal} {New. J.
  Phys.}\ }\textbf {\bibinfo {volume} {10}},\ \bibinfo {pages} {033036}
  (\bibinfo {year} {2008})}\BibitemShut {NoStop}%
\bibitem [{\citenamefont {Fortov}\ \emph {et~al.}(2005)\citenamefont {Fortov},
  \citenamefont {Ivlev}, \citenamefont {Khrapak}, \citenamefont {Khrapak},\
  and\ \citenamefont {Morfill}}]{fortov2005complex}%
  \BibitemOpen
  \bibfield  {author} {\bibinfo {author} {\bibfnamefont {V.~E.}\ \bibnamefont
  {Fortov}}, \bibinfo {author} {\bibfnamefont {A.~V.}\ \bibnamefont {Ivlev}},
  \bibinfo {author} {\bibfnamefont {S.~A.}\ \bibnamefont {Khrapak}}, \bibinfo
  {author} {\bibfnamefont {A.~G.}\ \bibnamefont {Khrapak}}, \ and\ \bibinfo
  {author} {\bibfnamefont {G.~E.}\ \bibnamefont {Morfill}},\ }\href@noop {}
  {\bibfield  {journal} {\bibinfo  {journal} {Phys. Rep.}\ }\textbf {\bibinfo
  {volume} {421}},\ \bibinfo {pages} {1} (\bibinfo {year} {2005})}\BibitemShut
  {NoStop}%
\bibitem [{\citenamefont {Khrapak}\ \emph {et~al.}(2002)\citenamefont
  {Khrapak}, \citenamefont {Ivlev}, \citenamefont {Morfill},\ and\
  \citenamefont {Thomas}}]{khrapak2002ion}%
  \BibitemOpen
  \bibfield  {author} {\bibinfo {author} {\bibfnamefont {S.~A.}\ \bibnamefont
  {Khrapak}}, \bibinfo {author} {\bibfnamefont {A.~V.}\ \bibnamefont {Ivlev}},
  \bibinfo {author} {\bibfnamefont {G.~E.}\ \bibnamefont {Morfill}}, \ and\
  \bibinfo {author} {\bibfnamefont {H.~M.}\ \bibnamefont {Thomas}},\
  }\href@noop {} {\bibfield  {journal} {\bibinfo  {journal} {Phys. Rev. E}\
  }\textbf {\bibinfo {volume} {66}},\ \bibinfo {pages} {046414} (\bibinfo
  {year} {2002})}\BibitemShut {NoStop}%
\bibitem [{\citenamefont {Khrapak}(2014)}]{khrapak2014accurate}%
  \BibitemOpen
  \bibfield  {author} {\bibinfo {author} {\bibfnamefont {S.~A.}\ \bibnamefont
  {Khrapak}},\ }\href@noop {} {\bibfield  {journal} {\bibinfo  {journal} {Phys.
  Plasmas}\ }\textbf {\bibinfo {volume} {21}},\ \bibinfo {pages} {044506}
  (\bibinfo {year} {2014})}\BibitemShut {NoStop}%
\end{thebibliography}
\providecommand{\noopsort}[1]{}\providecommand{\singleletter}[1]{#1}%

\end{document}